\documentclass[twocolumn,trackchanges]{aastex631}

\usepackage{graphicx}
\usepackage{subfigure}
\usepackage{paralist}
\usepackage{amssymb}
\usepackage{bm}
\usepackage{bbding}
\usepackage[fleqn]{amsmath}
\usepackage{multirow}
\hypersetup{linkcolor=red,citecolor=blue,filecolor=cyan,urlcolor=magenta}

\shorttitle{The ZLK effect inside MMRs}
\shortauthors{Lei et al.}

\graphicspath{{./}{figures/}}

\begin{document}

\title{The von Zeipel--Lidov--Kozai effect inside mean motion resonances with applications to trans-Neptunian objects}

\correspondingauthor{Hanlun Lei}
\email{leihl@nju.edu.cn}

\author{Hanlun Lei}
\author{Jian Li}
\author{Xiumin Huang}
\affiliation{School of Astronomy and Space Science, Nanjing University, Nanjing 210023, China}
\affiliation{Key Laboratory of Modern Astronomy and Astrophysics in Ministry of Education, Nanjing University, Nanjing 210023, China}
\author{Muzi Li}
\affiliation{Shanghai Aerospace control Technology Institude, Shanghai 201109, China}



\begin{abstract}
Secular dynamics inside MMRs plays an essential role in governing the dynamical structure of the trans-Neptunian region and sculpting the orbital distribution of trans-Neptunian objects (TNOs). In this study, semi-analytical developments are made to explore the von Zeipel--Lidov--Kozai (ZLK) resonance inside mean motion resonances (MMRs). To this end, a semi-secular model is formulated by averaging theory and then a single-degree-of-freedom integrable model is achieved based on the adiabatic invariance approximation. In particular, we introduce a modified adiabatic invariant, which is continuous around the separatrices of MMRs. During the long-term evolution, both the resonant Hamiltonian and the adiabatic invariant remain unchanged, thus phase portraits can be produced by plotting level curves of the adiabatic invariant with given Hamiltonian. The phase portraits provide global pictures to predict long-term behaviors of the eccentricity, inclination and argument of pericenter. Applications to some representative TNOs inside MMRs (2018 VO$_{137}$, 2005 SD$_{278}$, 2015 PD$_{312}$, Pluto, 2004 HA$_{79}$, 1996 TR$_{66}$ and 2014 SR$_{373}$) show good agreements between the numerically propagated trajectories under the full $N$-body model and the level curves arising in phase portraits. Interestingly, 2018 VO$_{137}$ and 2005 SD$_{278}$ exhibit switching behaviors during the long-term evolution and currently they are inside 2:5 MMR with Neptune.
\end{abstract}



\section{Introduction}
\label{Sect1}

According to the independent works performed by \citet{von1910application}, \citet{lidov1962evolution} and \citet{kozai1962secular}, the coupled oscillations between eccentricity and inclination of test particles in the long-term evolution are attributed to the von Zeipel--Lidov--Kozai (ZLK) effect \citep{ito2019lidov}. In the Kuiper belt, it is known that the ZLK effect inside mean motion resonances (MMRs) are important to understand the orbital distribution of trans-Neptunian objects (TNOs) \citep{gomes2005origin, gomes2011origin, gallardo2012survey, saillenfest2020long}. To understand the ZLK effect of TNOs inside MMRs with Neptune, it is of significance to formulate secular models \citep{saillenfest2016long}. After averaging out the short-period terms from the Hamiltonian, the resulting semi-secular model is of two degrees of freedom. One degree of freedom is associated with the MMR and the other degree of freedom is associated with the long-term evolution \citep{saillenfest2020long}. To study secular dynamics inside MMRs analytically, it is required to further reduce one degree of freedom from the semi-secular model. There are varieties of methods to reach this goal.

The simplest approach is to assume the critical argument associated with the MMR at the libration center \citep{kozai1985secular, yoshikawa1989survey, nesvorny2002perturbative, wan2007exploration, saillenfest2017study, li2021apsidal, Pons2022Secular}. By fixing the resonant angle or its amplitude to zero (this assumption corresponds to an adiabatic invariant equal to zero, as discussed latter), the degree of freedom associated with MMR disappears and the dynamical model immediately reduces to a one-degree-of-freedom integrable system. In the resulting reduced model, the motion of TNOs inside MMRs happens on the isolines of energy integral \citep{morbidelli2002modern}. Thus, phase portraits can be used to estimate the ranges of variations of orbital elements. Such an approximate model could provide a reasonable approximation for those particles located inside MMRs deeply. For a better approximation, some researchers assumed that the resonant angle associated with MMR varies along a sinusoid evolution with constant center, frequency and amplitude \citep{gomes2005origin, gomes2011origin, gallardo2012survey, brasil2014dynamical, huang2018kozai}. Under this assumption, the degree of freedom associated with MMR is decoupled from the other degree of freedom. Thus, it becomes possible to average the Hamiltonian over the period of the resonant angle and thus the degree of freedom associated with MMR is eliminated. However, in the original system the libration center, frequency and amplitudes associated with MMR are changed during the long-term evolution \citep{saillenfest2016long}.

In order to understand the chaotic and quasi-periodic libration zones on the representative plane \citep{wisdom1982origin, wisdom1983chaotic, murray1984structure}, \citet{wisdom1985perturbative} developed a semi-analytical perturbation theory for the secular dynamics near the 3:1 mean-motion resonance within the framework of planar elliptic restricted three-body problem (ERTBP). Averaging the original system over the period of the resonant angle leads to the evolutionary equations, which can be used to describe the long-term behaviors of the slow variable. Following the same idea, \citet{henrard1987perturbative} extended Wisdom's perturbation theory to the 2:1 Jovian resonance, and \citet{yokoyama1992application} applied Wisdom's perturbative method to asteroids inside 5:2 and 7:3 Jovian resonances within the framework of planar ERTBP. A generalization of Wisdom's perturbative method can be found in \citet{yokoyama1996simple}.

Based on adiabatic invariance approximation, \citet{henrard1986perturbation} developed a semi-numerical perturbation method and proposed a general numerical procedure allowing description of the dynamics associated with a two- or more-degree-of-freedom separable Hamiltonian. Especially, \citet{henrard1990semi} presented a numerical description for the action--angle variables of the separable Hamiltonian system. The Hamiltonian and the adiabatic invariant keep unchanged in the long-term evolution, meaning that the motion of resonant objects happens on the isolines of the Hamiltonian and adiabatic invariant. As a result, two types of phase portraits revealing global structures in the phase space can be produced: (a) plotting level curves of Hamiltonian with given adiabatic invariant \citep{morbidelli2002modern, saillenfest2016long, saillenfest2020long} and (b) plotting level curves of adiabatic invariant with given Hamiltonian \citep{wisdom1985perturbative, henrard1989motion, henrard1990semi, sidorenko2014quasi}. More discussions about perturbative treatments can be found in \citet{saillenfest2016long}, \citet{saillenfest2020long}, \citet{efimov2020analytically} and references therein.

The perturbation method based on adiabatic invariance approximation has been widely used in different contexts. \citet{saillenfest2016long} formulated semi-analytical one-degree-of-freedom integrable models for secular evolutions of minor bodies inside MMRs with Neptune. In this model, the precise variation of the resonant angle is taken into account to calculate the adiabatic invariant and the secular evolutions of TNOs are represented by level curves of Hamiltonian with given adiabatic invariant. \citet{saillenfest2017study} applied such a one-degree-of-freedom secular model to the distant trans-Neptunian region, showing pathways to high-perihelion distances and a ``trapping mechanism" maintaining the objects on distant orbits for billions of years. Review about the long-term dynamics of resonant TNOs can be found in \citet{gallardo2012survey} and \citet{saillenfest2020long}.

In this work, we adopt perturbative treatments based on adiabatic invariant approximation to deal with the ZLK resonance inside MMRs. Traditionally, the adiabatic invariant is defined as the oriented area enclosed by the isoline of Hamiltonian, see \citet{morbidelli2002modern} for example. As shown by \citet{saillenfest2016long} (see Section 5.2 in their work for a detailed discussion), the adiabatic invariant is not continuous at separatrix crossing. Thus, for those trajectories with switching behaviours, different phase portraits (level curves of Hamiltonian with given adiabatic invariant) should be matched in order to predict long-term behaviours. For a given TNO, it is difficult to provide the magnitudes of adiabatic invariant when it is inside and outside MMRs. To overcome the difficulty, we define an adiabatic invariant which is an extension of the standard definition and is continuous inside and outside MMRs. Then, phase portraits are produced by plotting level curves of adiabatic invariant with given Hamiltonian and. At last, the analytical developments are applied to known TNOs inside MMRs. The analytical models are validated by comparing analytical results with those of numerical integrations made under the full $N$-body model.

The remaining part of this article is organized as follows. In Section \ref{Sect2}, the basic dynamical models are briefly introduced, including the full $N$-body model, the simplified $N$-body model and the semi-secular model. The one-degree-of-freedom integrable model is developed based on the adiabatic invariance approximation and then it is validated by comparing the analytical results with those of numerical integrations made in Section \ref{Sect4}. Applications to real TNOs inside MMRs are provided in Section \ref{Sect5} and conclusions of this work are summarized in Section \ref{Sect6}.

\begin{figure*}
\centering
\includegraphics[width=0.45\textwidth]{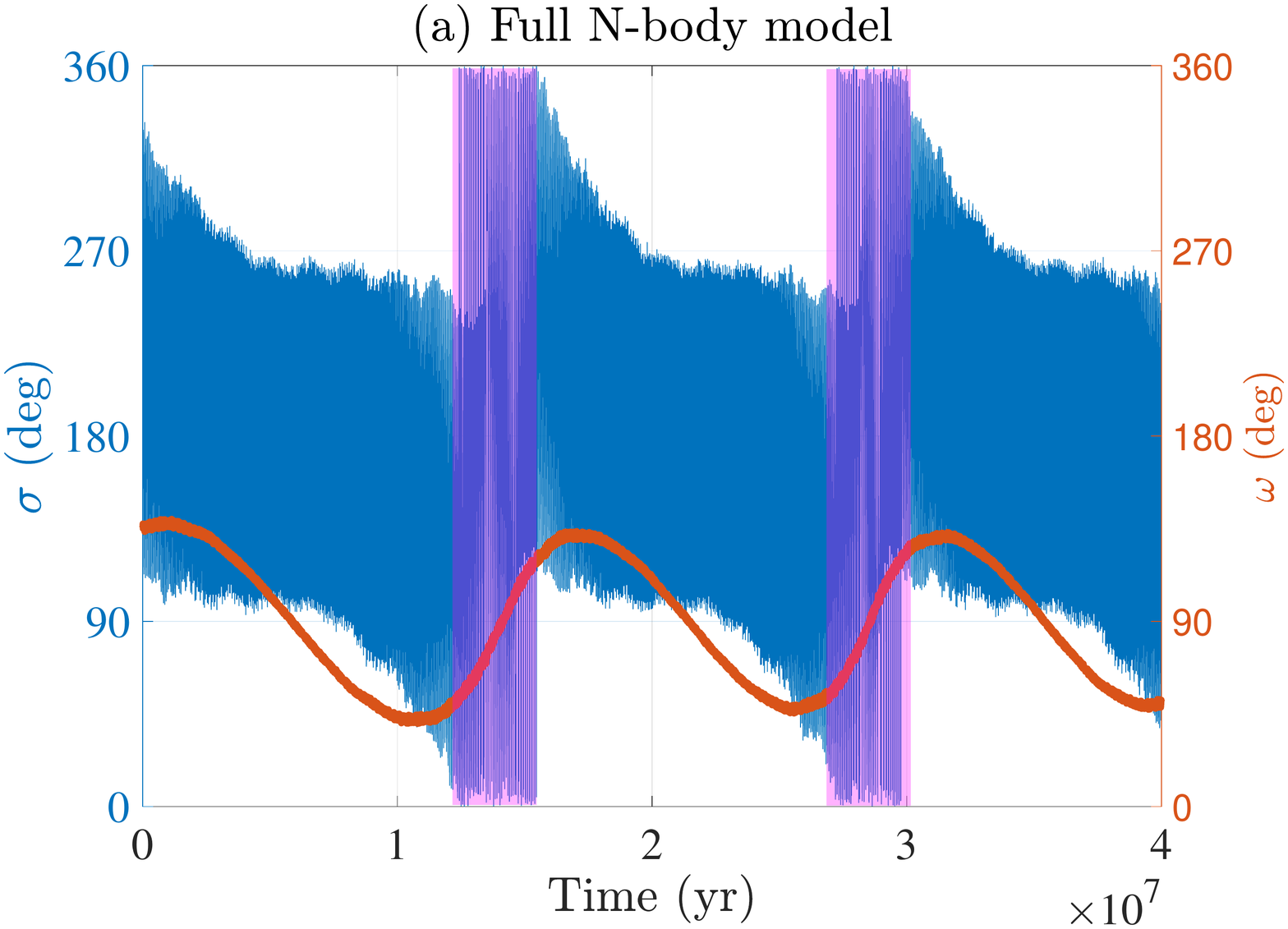}
\includegraphics[width=0.45\textwidth]{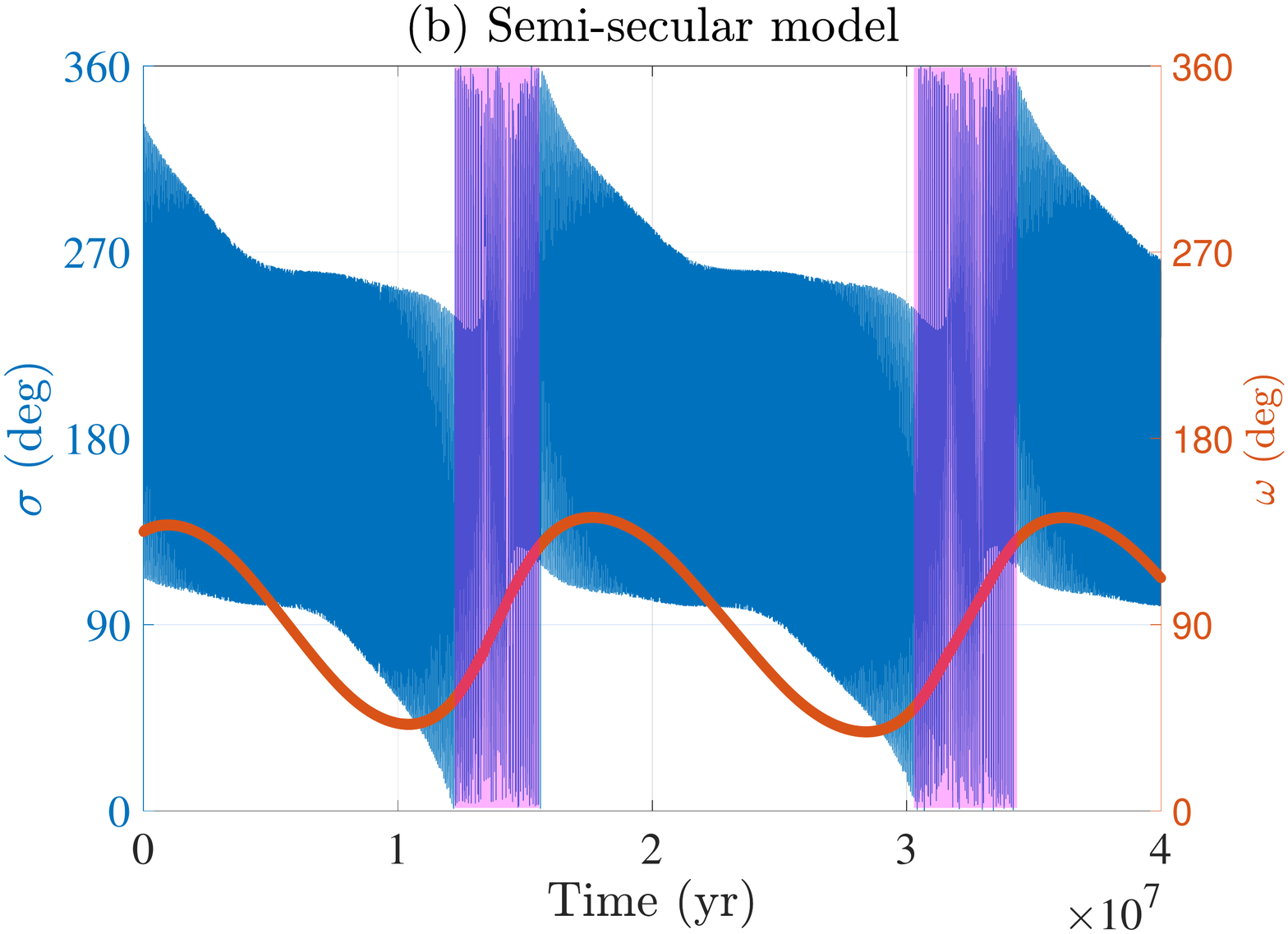}
\caption{Time histories of the critical argument $\sigma = 5\lambda - 2\lambda_8 -3\varpi$ and argument of perihelion $\omega$ for the orbits of 2018 VO$_{137}$ numerically propagated under the full $N$-body model (\emph{left panel}) and the semi-secular model (\emph{right panel}). The portions in circulation are marked in shade. The results produced in the semi-secular model agree well with the ones under the full $N$-body model. In simulations, the initial epoch is fixed on December 17th, 2020. Under the full $N$-body model, the initial state of 2018 VO$_{137}$ is provided by the osculating elements $a= 55.381$ $\rm{au}$, $e=0.184$, $I=39.0^{\circ}$, $\Omega= 42.7^{\circ}$, $\omega= 135.599997^{\circ}$ and $M=234.799997^{\circ}$ (please refer to https://minorplanetcenter.net for retrieving the elements). Under the semi-secular model, the initial state of 2018 VO$_{137}$ is provided by the associated averaged elements, which are given in Table \ref{Tab2}.
}
\label{Fig0}
\end{figure*}

\section{Dynamical models}
\label{Sect2}

Let us consider the outer Solar system model (the full N-body model), where the mass of terrestrial planets are added to the Sun and the mutual gravitational attractions between giant planets (Jupiter, Saturn, Uranus and Neptune) are taken into account. The trajectories propagated under such a full $N$-body model are taken as references to validate semi-secular and secular models formulated in this work.

Considering the fact that the eccentricities of giant planets are generally small and all the giant planets hold very low inclinations relative to the invariable plane, it is reasonable to assume that four giant planets are moving on circular and co-planar orbits around the Sun, i.e., giant planets are in the invariable plane of the outer Solar system. Under such an assumption, the Hamiltonian function that governs the evolution of small bodies involved in the system can be written as \citep{morbidelli1993secular}
\begin{equation}\label{Eq1}
{\cal H} =  - \frac{{{\mu _0}}}{{2a}} + \sum\limits_{i = 5}^8 {{{\dot \lambda}_i}{\Lambda _i}} - \sum\limits_{i = 5}^8 {{{\cal R}_i}},
\end{equation}
where $\mu_0 = {\cal G} m_0$ is the gravitational parameter of the Sun, $a$ is the semimajor axis of the test particle, ${\dot \lambda}_i$ is the time derivative of mean longitude for the $i$-th planet, and $\Lambda_i$ is the momentum conjugated to the mean longitude of the $i$-th planet $\lambda_i$. The subscripts 5,6,7 and 8 stand for Jupiter, Saturn, Uranus and Neptune, respectively. The planetary disturbing function ${\cal R}_i$ is given by
\begin{equation*}
{\cal R}_i = {{\mu _i}\left( {\frac{1}{{\left| {{\bm r} - {{\bm r}_i}} \right|}} - \frac{{{\bm r} \cdot {{\bm r}_i}}}{{{{\left| {{{\bm r}_i}} \right|}^3}}}} \right)},\quad i=5,6,7,8.
\end{equation*}
Due to the assumption that the orbits of giant planets are known and fixed, the dynamical model represented by equation (\ref{Eq1}) is a simplified N-body model, which is of 3 degrees of freedom.

The orbits of the test particle and the $i$-th giant planet around the Sun are described by orbital elements: the semimajor axis $a (a_i)$, the eccentricity $e (e_i)$, the inclination $I(I_i)$, the longitude of ascending node $\Omega(\Omega_i)$, the argument of pericenter $\omega(\omega_i)$ and mean anomaly $M(M_i)$. Usually, the longitude of pericenter $\varpi (\varpi_i)$ and mean longitude $\lambda (\lambda_i)$ are used for low-eccentricity and/or low-inclination cases. These variables are described in the Sun-centered right-handed inertial coordinate system with the invariable plane of the outer Solar system as the fundamental plane and the total angular momentum vector as the direction of the $z$-axis. In addition, the time and space variables are normalized by taking the mean semimajor axis between the Sun and Neptune (over $2\times 10^{7}$ yrs) as the unit of length, total mass of the Sun and Neptune as the unit of mass and the two-body orbital period of Neptune divided by $2\pi$ as unit of time. The physical parameters adopted in this work are given in Table \ref{Tab1}.

\begin{table}
\small
\centering
\caption{Physical parameters adopted in this study. The Julian year (yr) consists of 365.25 Julian days and $m_E$ is the mass of the Earth. The variables $a_i (i=5,6,7,8)$ are the mean semimajor axes of Jupiter, Saturn, Uranus and Neptune. See the text for the computation of $a_i (i=5,6,7,8)$ and ${\dot \lambda}_8$.}
\begin{tabular*}{\hsize}{@{}@{\extracolsep{\fill}}lc@{}}
\hline
System of units & Magnitude\\
\hline\hline
Unit of time (yr) & 26.2948188764\\
Unit of length (au) & 30.1094546076\\
Unit of mass ($m_E$) & 332965.1922595720\\
\hline
Dimensionless parameters & Magnitude\\
\hline\hline
$\mu_0$ & 9.9994848907E-1\\
$\mu_5$ & 9.5473704974E-4\\
$\mu_6$ & 2.8586954577E-4\\
$\mu_7$ & 4.3659930385E-5\\
$\mu_8$ & $1.0-\mu_0$\\
\hline
$a_5$ & 0.1727862090\\
$a_6$ & 0.3173329830\\
$a_7$ & 0.6382793627\\
$a_8$ & 1.0\\
\hline
${\dot \lambda}_8$     & 1.0027387100\\
$T_8$ (yr)& 164.7639788626\\
\hline
\end{tabular*}
\label{Tab1}
\end{table}

To formulate the Hamiltonian model, we introduce the modified Delaunay variables as follows \citep{morbidelli2002modern}:
\begin{equation}\label{Eq4}
\begin{aligned}
\Lambda  &= \sqrt {{\mu _0}a} ,\quad\quad \lambda  = M + \varpi, \\
P &= \sqrt {{\mu _0}a} \left( {1 - \sqrt {1 - {e^2}} } \right),\quad p =  - \varpi, \\
Q &= \sqrt {{\mu _0}a\left( {1 - {e^2}} \right)} \left( {1 - \cos i} \right),\quad q =  - \Omega, \\
{\Lambda _i} &, \quad\quad\quad {\lambda _i}.
\end{aligned}
\end{equation}
Using Delaunay variables, the Hamiltonian of dynamical system can be written as follows:
\begin{equation}\label{Eq5}
{\cal H} =  - \frac{{\mu _0^2}}{{2{\Lambda ^2}}} + \sum\limits_{i = 5}^8 {{{\dot \lambda}_i}{\Lambda _i}}  - \sum\limits_{i = 5}^8 {{{\cal R}_i}}.
\end{equation}

For an object located inside $k_p$:$k$ resonance with Neptune, the resonant angle is usually given by
\begin{equation}\label{Eq6}
\sigma  = k\lambda  - {k_p}{\lambda _8} + \left( {{k_p} - k} \right)\varpi
\end{equation}
which is called the eccentricity-type resonant argument \citep{murray1999solar}. Usually, the integer $\left|k - k_p\right|$ is called the resonance order.

To formulate the semi-secular model, the following canonical variables are introduced,
\begin{equation}\label{Eq7}
\begin{aligned}
\Sigma  &= \frac{1}{k}\Lambda ,\quad \sigma  = k\lambda  - {k_p}{\lambda _8} - \left( {{k_p} - k} \right)p,\\
U &=  - P - \frac{{{k_p} - k}}{k}\Lambda ,\quad u = q - p,\\
V &=  - P - Q - \frac{{{k_p} - k}}{k}\Lambda ,\quad v =  - q,\\
W &= {\Lambda _8} + \frac{{{k_p}}}{k}\Lambda ,\quad w = {\lambda _8}
\end{aligned}
\end{equation}
with the generating function
\begin{equation*}
{\cal S} = k\lambda \Sigma  - p\left( {U + {k_p}\Sigma  - k\Sigma } \right){\rm{ + }}q\left( {U - V} \right){\rm{ + }}{\lambda _8}\left( {W - {k_p}\Sigma } \right).
\end{equation*}
In terms of the classical orbital elements, the canonical variables given by equation (\ref{Eq7}) can be expressed as
\begin{equation}\label{Eq8}
\begin{aligned}
\Sigma  &= \frac{1}{k}\sqrt {{\mu _0}a} ,\quad\quad \sigma  = k\lambda  - {k_p}{\lambda _8} + \left( {{k_p} - k} \right)\varpi, \\
U &= \sqrt {{\mu _0}a} \left( {\sqrt {1 - {e^2}}  - \frac{{{k_p}}}{k}} \right),\quad u = \omega, \\
V &= \sqrt {{\mu _0}a} \left( {\sqrt {1 - {e^2}} \cos I - \frac{{{k_p}}}{k}} \right),\quad v = \Omega, \\
W &= {\Lambda _8} + \frac{{{k_p}}}{k}\sqrt {{\mu _0}a} ,\quad\quad w = {\lambda _8}.
\end{aligned}
\end{equation}

Among the angular variables shown in the Hamiltonian, it is not difficult to see that the mean longitudes ($\lambda$ and $\lambda_i$) are short-period variables, $\sigma$ is a semi-secular periodic variable and $\omega$ and $\Omega$ are long-period variables \citep{saillenfest2020long}. Based on this fact, those terms related to the short-period variables produce short-period influences upon the secular dynamics, thus they can be averaged out from the Hamiltonian by means of perturbation theory in order to study long-term evolutions \citep{hori1966theory, deprit1969canonical}.

Concerning the disturbing functions ${\cal R}_{5,6,7}$ (non-resonant configurations), there are two short-period variables: $\lambda$ and $\lambda_{i}$ ($i=5,6,7$). Thus, it requires to perform double averages over orbital periods of the test particle and giant planets to remove the short-period influences by \citep{morbidelli1993secular, thomas1996kozai}
\begin{equation}\label{Eq10}
\begin{aligned}
{\cal R}_i^* &= \frac{{\mu _i}}{{4{\pi ^2}}}\int\limits_0^{2\pi } {\int\limits_0^{2\pi } {\left( {\frac{1}{{\left| {{\bm r} - {{\bm r}_i}} \right|}} - \frac{{{\bm r} \cdot {{\bm r}_i}}}{{{{\left| {{{\bm r}_i}} \right|}^3}}}} \right){\rm d}\lambda {\rm d}{\lambda _i}} }\\
& = \frac{{\mu _i}}{{4{\pi ^2}}}\int\limits_0^{2\pi } {\int\limits_0^{2\pi } {{\frac{1}{{\left| {{\bm r} - {{\bm r}_i}} \right|}}}{\rm d}\lambda {\rm d}{\lambda _i}} }, \quad i=5,6,7.
\end{aligned}
\end{equation}
In practical simulations, the double-averaged disturbing functions given by Eq. (\ref{Eq10}) can be truncated at the fourth order in semimajor axis ratio $a_i/a$ and they can be explicitly expressed by \citep{saillenfest2020long}
\begin{equation}\label{Eq11}
\begin{aligned}
{\cal R}_i^* &= \frac{{{\mu _i}}}{a} + \frac{1}{8}\frac{{{\mu _i}}}{a}{\left( {\frac{{{a_i}}}{a}} \right)^2}\frac{1}{{{{\left( {1 - {e^2}} \right)}^{3/2}}}}\left( {3{{\cos }^2}I - 1} \right)\\
& + \frac{9}{{1024}}\frac{{{\mu _i}}}{a}{\left( {\frac{{{a_i}}}{a}} \right)^4}\frac{1}{{{{\left( {1 - {e^2}} \right)}^{7/2}}}}{\cal F}\left( {e,I,\omega } \right),\quad i=5,6,7
\end{aligned}
\end{equation}
where
\begin{equation*}
\begin{aligned}
{\cal F}\left( {e,I,\omega } \right) &= \left( {3 - 30{{\cos }^2}I + 35{{\cos }^4}I} \right)\left( {2 + 3{e^2}} \right)\\
&+ 10\left( {7{{\cos }^2}I - 1} \right){e^2}{\sin ^2}I\cos \left( {2\omega } \right) .
\end{aligned}
\end{equation*}

About the disturbing function ${\cal R}_8$ (resonant configuration), there is only one short-period variable $\lambda_8$. Thus, it requires to perform a single average for ${\cal R}_8$ over $k$ times Neptune's orbital period as follows \citep{gallardo2006atlas}:
\begin{equation}\label{Eq9}
{\cal R}_8^{^{{*}}} = \frac{{\mu _8}}{{2k\pi }}\int\limits_0^{2k\pi } {\left( {\frac{1}{{\left| {{\bm r} - {{\bm r}_8}} \right|}} - \frac{{{\bm r} \cdot {{\bm r}_8}}}{{{{\left| {{{\bm r}_8}} \right|}^3}}}} \right){\rm d}{\lambda _8}},
\end{equation}
which can be calculated by direct numerical quadrature \citep{schubart1968long}. Evaluating equation (\ref{Eq9}) by numerical quadrature is applicable for test particles with arbitrary semimajor axes, eccentricities and inclinations.

To be consistent, the orbital elements appearing in ${\cal R}_i^* (i=5,6,7,8)$ should be replaced by the canonical variables $(\sigma, \Sigma, u, U)$. Removing those constant terms associated with $\Lambda_{5,6,7}$ from the Hamiltonian (without changing the Hamiltonian dynamics), we can obtain the averaged Hamiltonian function without short-period terms, given by
\begin{equation}\label{Eq12}
{{\cal H}^*} = - \frac{{\mu _0^2}}{{2{{\left( {k\Sigma } \right)}^2}}} - {k_p} {{\dot \lambda}_8}\Sigma  - \sum\limits_{i = 5}^8 {{\cal R}_i^*\left(\sigma, \Sigma, u, U\right)}.
\end{equation}
In order to produce the value of ${\dot \lambda}_8  (= {\dot M}_8 + {\dot \varpi}_8)$, we numerically integrate the equations of motion under the full $N$-body model (or the OSS model) over $2 \times 10^7$ years. Then, we identify the frequency of Neptune's mean longitude ${\dot \lambda}_8$ by means of linear fitting for the time series function $\lambda_8 (t)$ and determine the mean semimajor axes of giant planets by numerically averaging the time series function $a_i (t)$. It should be mentioned that under the full $N$-body model the time evolution of the mean longitude of Neptune is not equal to mean $n_8$ (Neptune's mean motion) because there is a contribution of the time evolution of $\varpi_8$. Also, under the full $N$-body model, the mean semimajor axis $a_8$ of Neptune will not satisfy the Keplerian relationship with mean $n_8$.

The practical values of $a_i(i =5,6,7,8)$ and ${\dot \lambda}_8$ adopted in this study are provided in Table \ref{Tab1}. In this work, the frequency of Neptune's mean longitude is $\dot \lambda_8 = 1.00273871$ in normalized units, which is very close to the value given by \citet{murray1999solar} (it is $\dot \lambda_8 =1.0025723684$, as shown by Table A.3 in their textbook). The minor difference of $\dot \lambda_8$ in magnitude is due to the difference of dynamical models.

To make the magnitude of Hamiltonian be close to unity, we add some constant terms to the Hamiltonian without changing the Hamiltonian dynamics and then normalize it by the gravitational parameter of Neptune in the following manner (for convenience, we still use $\cal H$ to stand for the normalized Hamiltonian and it is called the resonant Hamiltonian from now on)
\begin{equation}\label{Eq13}
{\cal H} = \frac{1}{{{\mu _8}}}\left[ {{{\cal H}^{\rm{*}}} + \frac{{\mu _0^2}}{{2{{\left( {k{\Sigma _0}} \right)}^2}}} + {{\dot \lambda}_8}{k_p}{\Sigma _0} + \sum\limits_{i = 5}^7 {\frac{{{\mu _i}}}{{{a_0}}}} } \right],
\end{equation}
where $a_0$ and $\Sigma_0$ are the reference values of $a$ and $\Sigma$, given by ${\left( {\frac{{{k_p}}}{k}} \right)^2}a_0^3 = {\mu _0}$ and ${\Sigma _0} = \frac{1}{k}\sqrt {{\mu _0}{a_0}}$. It is noted that the dynamical models represented by equations (\ref{Eq12}) and (\ref{Eq13}) remain the same except that the time unit is magnified $\frac{1}{\mu_8}$ times in the latter model.

Evidently, the semi-secular model represented by equation (\ref{Eq13}) is of two degrees of freedom with $\sigma$ and $u$ as the angular coordinates. Since the angular variable $v = \Omega$ disappears from the resonant Hamiltonian, its conjugate momentum becomes an integral of motion, given by
\begin{equation}\label{Eq14}
V = \sqrt {\mu_0 a} \left( {\sqrt {1 - {e^2}} \cos I - \frac{{{k_p}}}{k}} \right) = {\rm const}.
\end{equation}
Assuming the semimajor axis at the nominal location of resonance, i.e. $a={a_c} = {\left( {\frac{{{\mu _0}}}{{n_8^2}}\frac{{{k^2}}}{{k_p^2}}} \right)^{1/3}}$, the integral of motion can be expressed by
\begin{equation}\label{Eq15}
V = \sqrt {\mu_0 {a_c}} \left( {\sqrt {1 - {{\tilde e}^2}} \cos \tilde I - \frac{{{k_p}}}{k}} \right) = {\rm const},
\end{equation}
where ${\tilde e}$ and $\tilde I$ are called the equivalent eccentricity and inclination.

Furthermore, we assume the equivalent eccentricity ${\tilde e}$ as zero and then it is possible to get the maximum inclination $I_{\max}$ standing for the magnitude of $V$ in the following manner:
\begin{equation}\label{Eq16}
V = \sqrt {\mu_0 {a_c}} \left( {\cos {I_{\max }} - \frac{{{k_p}}}{k}} \right) = {\rm const},
\end{equation}
The parameter $I_{\max}$ is used to specify the motion integral $V$ and it is similar to the Kozai parameter $i_0$ \citep{kozai1962secular}. Up to now, a one-to-one correspondence between $V$ and $I_{\max}$ has been made. As a result, the resonant Hamiltonian can be expressed as ${\cal H}\left(I_{\max}; \sigma, \Sigma, u, U\right)$, which determines a two-degree-of-freedom dynamical model, depending on the motion integral $V$ (or the equivalent parameter $I_{\max}$).

The Hamiltonian canonical relations lead to the equations of motion for trans-Neptunian objects inside MMRs as follows:
\begin{equation}\label{Eq17}
\begin{aligned}
\dot \sigma  =& \frac{{\partial {\cal H}}}{{\partial \Sigma }},\quad \dot \Sigma  =  - \frac{{\partial {\cal H}}}{{\partial \sigma }},\\
\dot u =& \frac{{\partial {\cal H}}}{{\partial U}},\quad \dot U =  - \frac{{\partial {\cal H}}}{{\partial u}},
\end{aligned}
\end{equation}
which are the equations of motion of the semi-secular model. The degree of freedom associated with the MMR is represented by $(\sigma, \Sigma)$ and the one associated with long term orbital evolution is represented by $(u, U)$. In the second degree of freedom, the angular variable is defined by $u = \omega$ and the action variable $U$ can be expressed by the equivalent eccentricity as $U = \sqrt {{\mu _0}{a_c}} \left( {\sqrt {1 - {{\tilde e}^2}}  - \frac{{{k_p}}}{k}} \right)$ where $a_c$ is the nominal location of $k_p$:$k$ MMR. Thus, the second degree of freedom can be represented by $(\omega, \tilde{e})$. Naturally, the resonant Hamiltonian can be denoted by ${\cal H}\left(I_{\max}; \sigma, \Sigma, \omega, \tilde{e}\right)$.

To validate the semi-secular model represented by equation (\ref{Eq17}), we take 2018 VO$_{137}$ as an example. It is currently located inside 2:5 resonance with Neptune, and its resonant argument is defined by $\sigma = 5\lambda - 2\lambda_8 - 3\varpi$. We propagate the orbits of asteroid 2018 VO$_{137}$ under both the full $N$-body model and the semi-secular model over $4.0 \times 10^{7}$ $\rm{yr}$. The time histories of the resonant argument $\sigma$ and argument of perihelion $\omega$ are reported in Fig. \ref{Fig0}.

From Fig. \ref{Fig0}, it is observed that, under the full $N$-body model, 2018 VO$_{137}$ transits between libration regions and circulation regions periodically and it is currently inside ZLK resonance with $\omega$ librating around $90^{\circ}$.

Comparing the evolutions in different models, we can see that the angles $\sigma$ and $\omega$ produced under the semi-secular model are in good agreement with the ones produced under the full $N$-body model. In the following sections, we will make analytical developments based on the semi-secular model.

\section{Secular models for resonant objects}
\label{Sect4}

As stated in the previous section, the semi-secular dynamical model for describing the long-term dynamics inside or around MMRs is of two degrees of freedom. The frequencies between the two degrees of freedom separate faraway from each other, i.e., $\left|\dot \sigma\right|  \gg \left|\dot \omega\right|$, meaning that such a dynamical model can be divided into `fast' and `slow' subsystems \citep{saillenfest2016long, saillenfest2020long}. Perturbation theory is powerful to deal with such a kind of separable Hamiltonian model \citep{wisdom1985perturbative, henrard1990semi}.

\subsection{Semi-analytical developments}
\label{Sect4-1}

Considering the separability of frequencies between the two degrees of freedom, let us freeze the variables of the `slow' subsystem during the timescale of the `fast' subsystem associated with $(\sigma, \Sigma)$ and denote the associated Hamiltonian by
\begin{equation}\label{Eq17_1}
{\cal H} (I_{\max}, \omega, \tilde{e}; \sigma, \Sigma)
\end{equation}
where the slow variables $(\omega, \tilde{e})$ are treated as parameters \citep{saillenfest2016long}. Obviously, the Hamiltonian model represented by equation (\ref{Eq17_1}) is of a single degree of freedom. According to the usual perturbative treatments, Arnold action-angle variables can be introduced for the `fast' degree of freedom \citep{henrard1990semi, morbidelli2002modern}:
\begin{equation}\label{Eq18}
{\psi _\sigma } = \frac{{2\pi }}{{{T_{\sigma} }}}t,\quad {J_\sigma } = \frac{1}{{2\pi }}\oint {\Sigma {\rm d}\sigma },
\end{equation}
where $T_\sigma$ is the libration period of $\sigma$. In geometry, ${2\pi J_\sigma }$ defined by the path integration in equation (\ref{Eq18}) stands for the signed area enclosed by the solution curve of the `fast' subsystem. See Fig. 4.2 in \citet{morbidelli2002modern} for the geometrical definition of ${2\pi J_\sigma }$. Using the action-angle variables, the Hamiltonian can be denoted by
\begin{equation}\label{Eq18_1}
{\cal H} (I_{\max}, \omega, \tilde{e}; J_{\sigma}),
\end{equation}
which determines a single-degree-of-freedom dynamical model for the `slow' subsystem with $J_{\sigma}$ as the motion integral. This approximation is a `adiabatic' process, and the motion integral $J_{\sigma}$ is referred to as an adiabatic invariant \citep{neishtadt1987change}.

In particular, when the motion integral ${J_\sigma }$ is provided, the trajectories in the dynamical model determined by equation (\ref{Eq18_1}) are level curves of resonant Hamiltonian in the space $(\omega, \tilde{e})$. Thus, the conventional phase portraits can be produced by plotting level curves of resonant Hamiltonian in the $(\omega, \tilde{e})$ space with given ${J_\sigma }$ \citep{morbidelli2002modern}. By analyzing the structures arising in the phase portraits, it is possible to know the global dynamics in the phase space. \citet{saillenfest2016long} pointed out that the computation of the Hamiltonian as well as its partial derivatives is time consuming.

There are several difficulties for the problem at hand. At first, the motion integral $J_\sigma$ is not continuous when particles switch between libration and circulation regions, so that the phase portraits need to be separately considered and then matched for phase spaces inside and outside MMRs (see for instance \citet{saillenfest2016long} for more details). Thus, it is inconvenient to describe those particles with switching behaviors between circulation and libration. Secondly, it is not an easy task to inversely solve the state $(\sigma, \Sigma, \omega, \tilde{e})$ for a given motion integral $J_\sigma$ because $J_\sigma$ is not an explicit function of the variables ($\sigma$, $\Sigma$, $\omega$, $\tilde{e}$). About the first point, \citet{neishtadt2004wisdom} and \citet{henrard1986perturbation} have made some attempts.

To make the adiabatic invariant be continuous around separatrices of the `fast' subsystem, we introduce the absolute of the area bounded by the isolines of resonant Hamiltonian
\begin{equation}\label{Eq19}
S\left( {{I_{\max }},{\cal H};\omega ,\tilde e} \right) = \left| {\oint {\Sigma {\rm d}\sigma } } \right|
\end{equation}
inside the libration zone and
\begin{equation}\label{Eq20}
S\left( {{I_{\max }},{\cal H};\omega ,\tilde e} \right) =  {\int\limits_0^{2\pi } {\left( {{\Sigma _{\rm up}} - {\Sigma _{\rm down}}} \right) {\rm d}\sigma } }
\end{equation}
inside the circulation zone as new adiabatic invariant. In equation (\ref{Eq20}), $\Sigma _{\rm up}$ is evaluated at the upper isoline of Hamiltonian and $\Sigma _{\rm down}$ is evaluated at the bottom isoline of Hamiltonian. Using the new adiabatic invariant $S$, the Hamiltonian can be written as
\begin{equation}\label{Eq20_1}
{\cal H} (I_{\max}, S; \omega, \tilde{e}),
\end{equation}
which is a single-degree-of-freedom dynamical model with $S$ as the motion integral (adiabatic invariant).

\begin{figure*}
\centering
\includegraphics[width=0.4\textwidth]{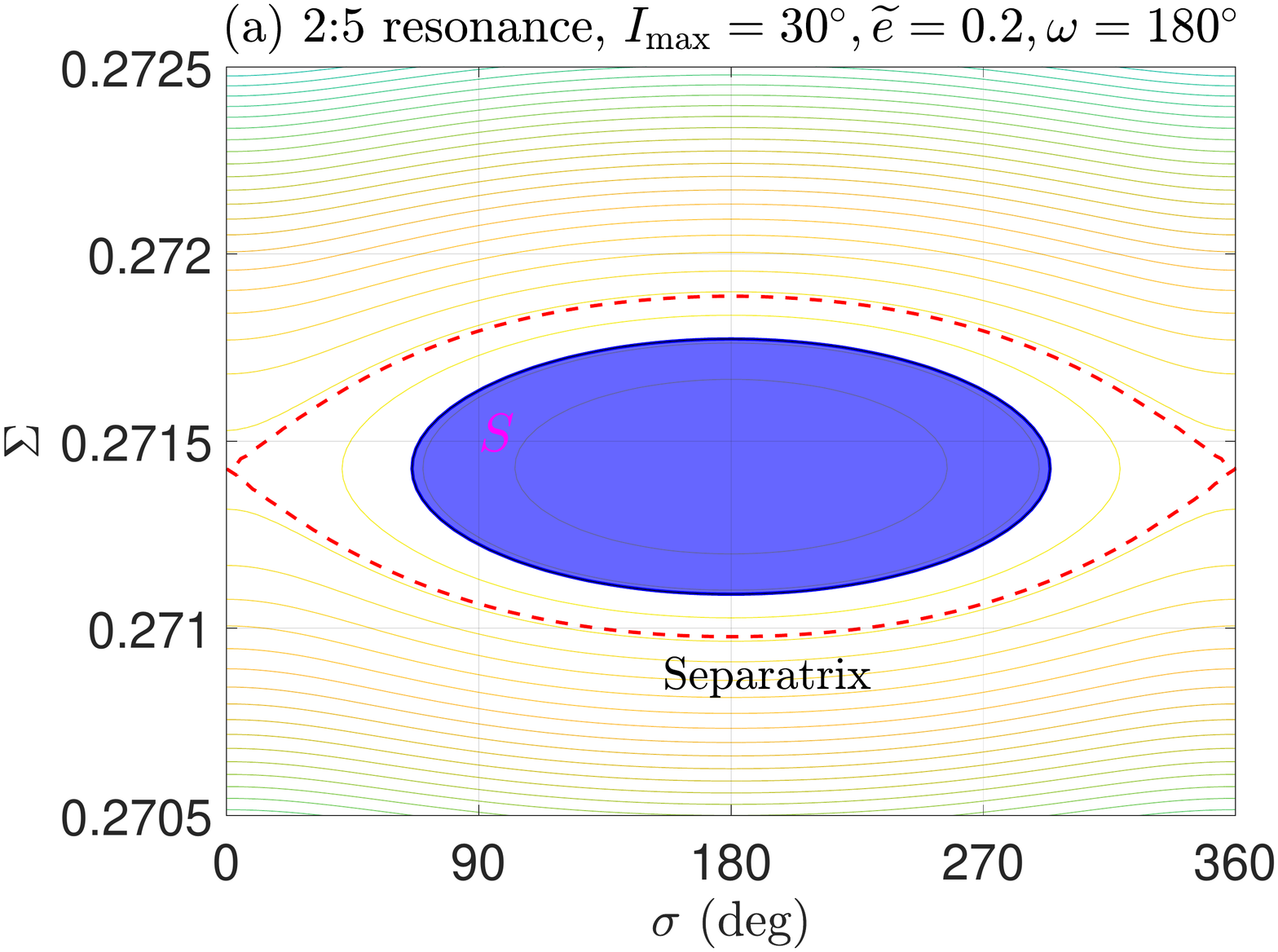}
\includegraphics[width=0.4\textwidth]{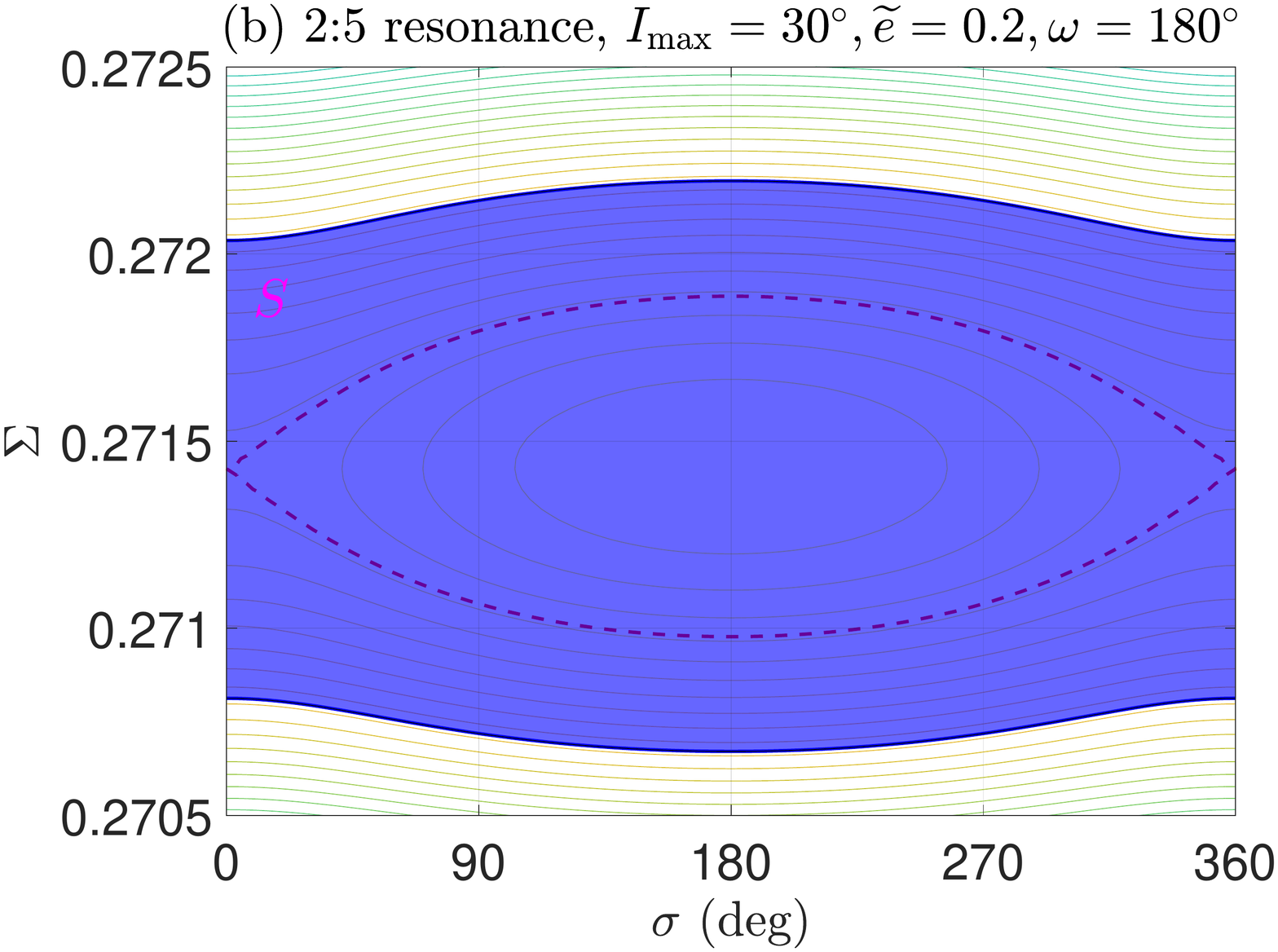}
\caption{Definition of the adiabatic invariant, $S$, corresponding to the area surrounded by isolines of resonant Hamiltonian in the phase space $(\sigma, \Sigma)$ for the case of 2:5 resonance (holding symmetric libration centers). The left panel is for the resonant case (i.e., ${\cal H} > {\cal H}_U$) and the right panel is for the circulation case (i.e., ${\cal H} < {\cal H}_U$). In both panels, the dynamical separatrices with ${\cal H} = {\cal H}_U$ are shown by red dashed lines.}
\label{Fig4}
\end{figure*}

\begin{figure*}
\centering
\includegraphics[width=0.4\textwidth]{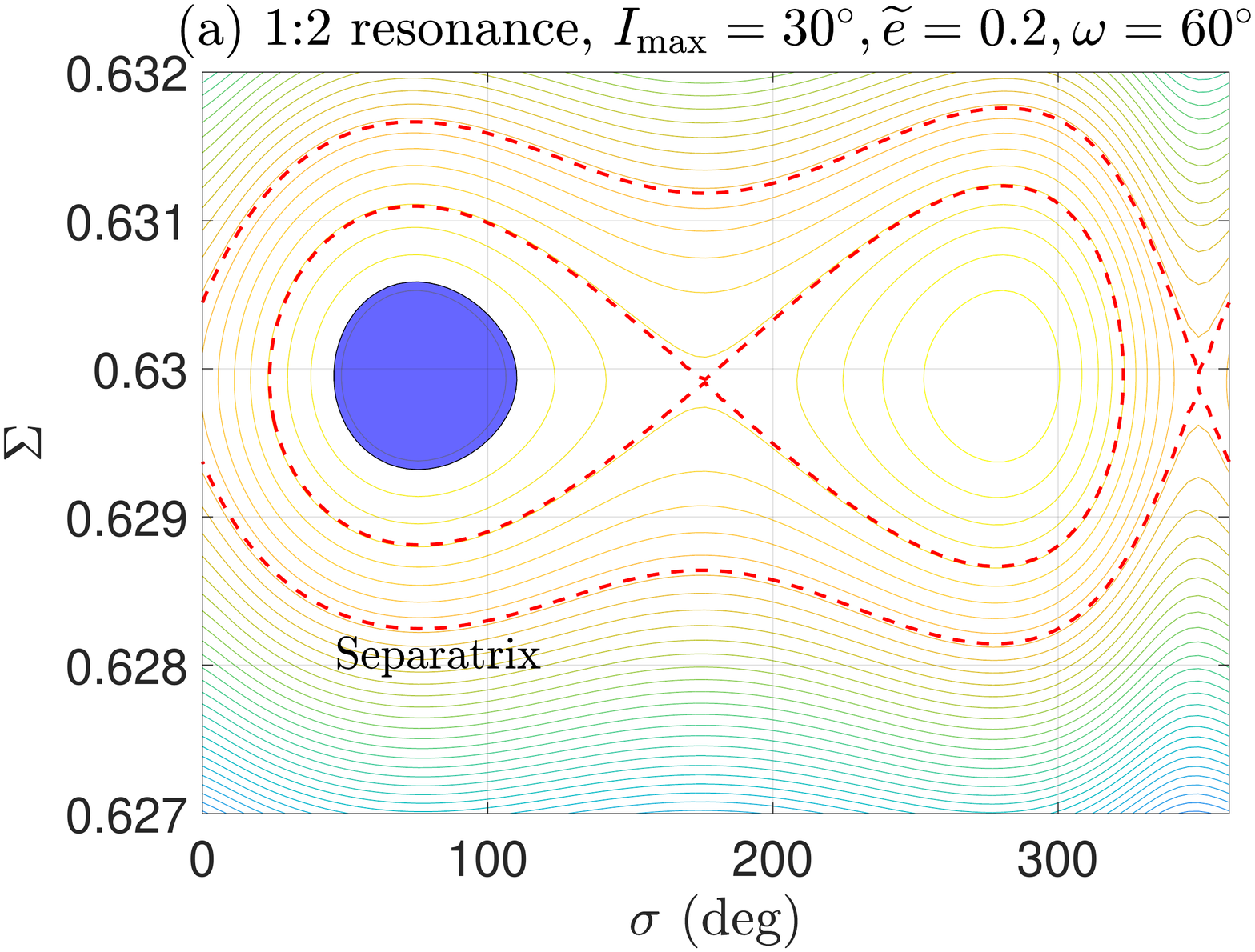}
\includegraphics[width=0.4\textwidth]{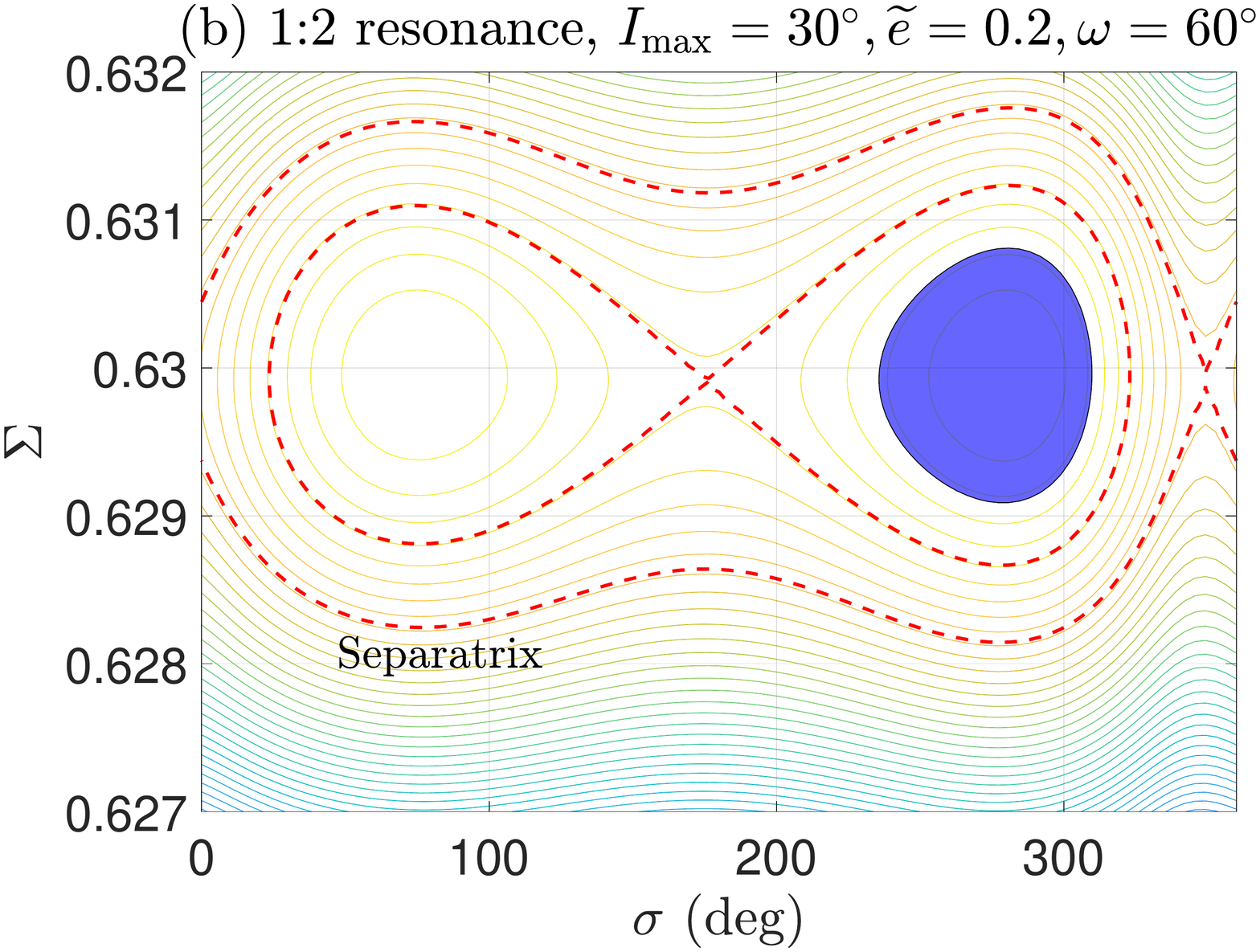}\\
\includegraphics[width=0.4\textwidth]{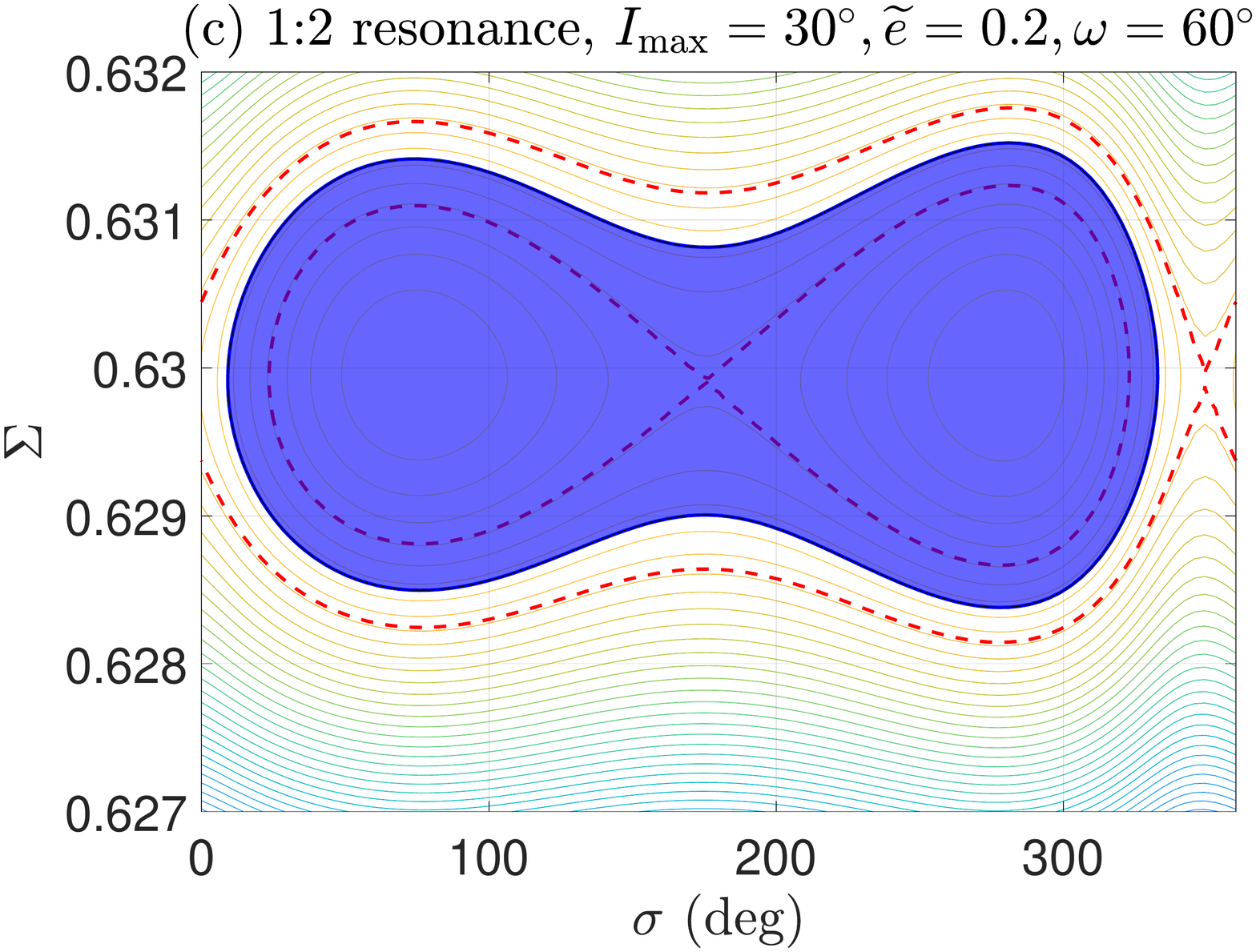}
\includegraphics[width=0.4\textwidth]{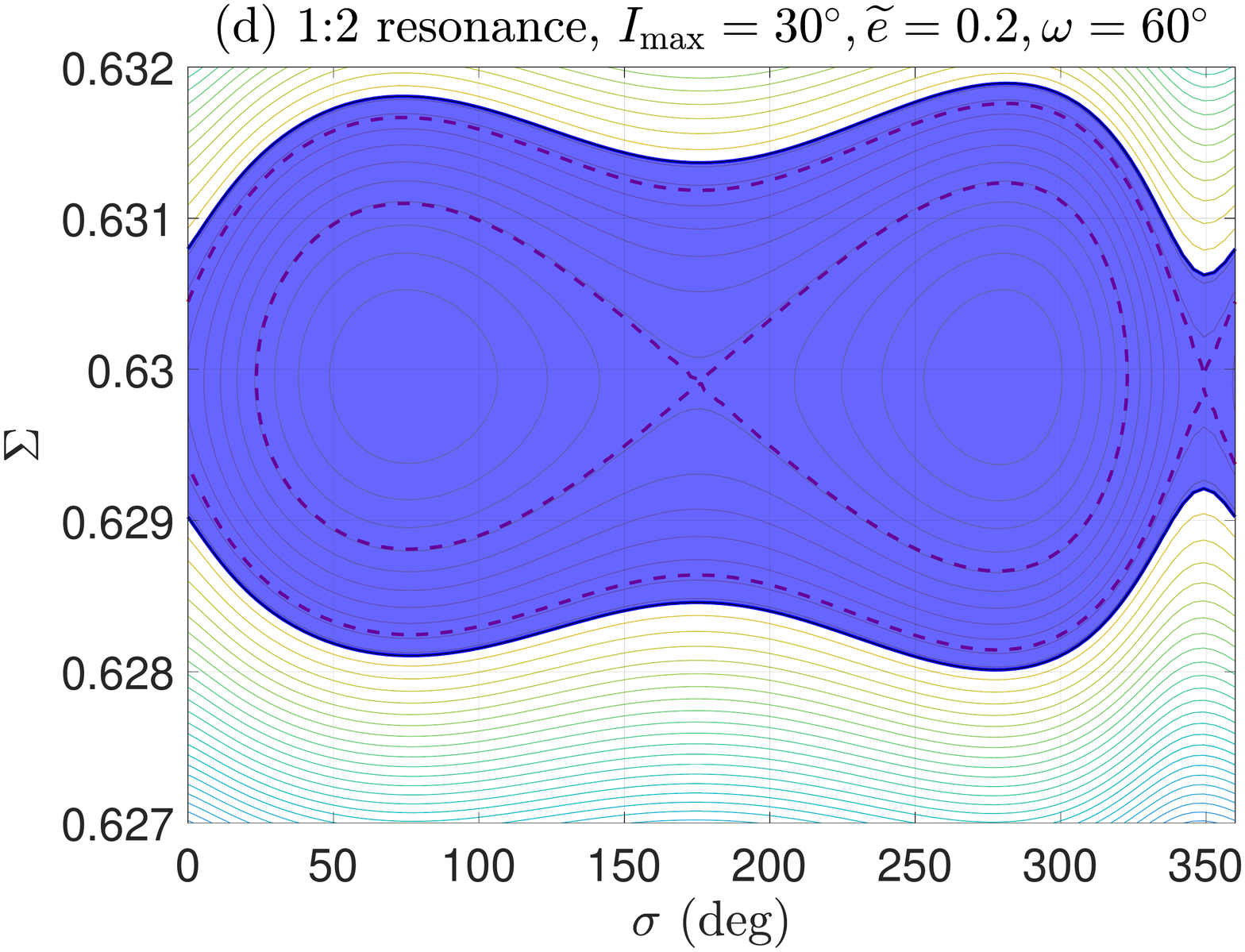}
\caption{Definition of the adiabatic invariant, $S$, represented by the area surrounded by isolines of resonant Hamiltonian in the phase space $(\sigma, \Sigma)$ for the case of 1:2 resonance (holding asymmetric libration centers). There are inner and outer separatrices, which are marked by red dashed lines. The Hamiltonian of the inner separatrix is denoted by ${\cal H}_U^{(1)}$ and that of the outer separatrix is denoted by ${\cal H}_U^{(2)}$. The upper panels are for the libration cases with tadpole-type trajectories (i.e., ${\cal H} > {\cal H}_U^{(1)}$), and the bottom panels are for the libration case with horseshoe-type trajectories (i.e., ${\cal H}_U^{(1)} > {\cal H} > {\cal H}_U^{(2)}$) and for the circulation case (${\cal H} < {\cal H}_U^{(2)}$).}
\label{Fig5}
\end{figure*}

Inside the libration zone, it is known that the isoline of resonant Hamiltonian is enclosed and isomorphic to a circle and thus in this case the area $S$ is exactly equivalent to the conventional motion integral (in magnitude it holds $S = 2\pi \left|J_{\sigma}\right|$). However, in the circulation zone, there are two isolines of resonant Hamiltonian in the phase space of MMR (see Figs \ref{Fig4} and \ref{Fig5} for details), one is located above the libration island and the other one is below the libration island. The definition given by equation (\ref{Eq20}) shows that $S$ stands for the area bounded by the top and bottom isolines. Evidently, the definition $S$ in the circulation case is different from the conventional one.

In summary, for the two degree of freedom dynamical model at hand, there are two conserved quantities including the resonant Hamiltonian ${\cal H}$ and the area $S$ enclosed by the isolines of resonant Hamiltonian. These two integrals provide two nonlinear constraints imbedded in the full four-dimensional phase space $(\sigma, \Sigma, \omega, \tilde{e})$. Thus, when ${\cal H}$ and $S$ are given, the motion of TNOs in the long-term evolution happens in a two-dimensional manifold. Such a two-dimensional manifold can be graphically illustrated by means of phase portraits \citep{henrard1989motion, henrard1990semi}, Poincar\'e surfaces of section \citep{yokoyama1996simple} and a set of guiding trajectories \citep{wisdom1985perturbative}.

According to \citet{lei2019three} and \citet{gallardo2020three}, the location of libration center is dependent on the eccentricity, inclination and argument of pericenter. Particularly, the 1:n-type MMRs have two stable libration centers in the phase space (meaning that horseshoe and horseshoe trajectories are possible), while the non-1:n type MMRs have only one stable libration center. In the following, the cases of 1:n and non-1:n MMRs are discussed separately.

In Fig. \ref{Fig4}, we show the geometrical definition of $S$ for the resonances of non-1:n type. In practice, we take 2:5 resonance with Neptune as an example. The dynamical separatrices arising in the phase portraits of MMR are shown by red dashed lines. In the case of 2:5 resonance, the region bounded by the isoline of given Hamiltonian is marked by shaded area and, in particular, the left panel of Fig. \ref{Fig4} shows the definition of $S$ inside the libration zone of MMR and the right panel of Fig. \ref{Fig4} shows the definition of $S$ inside the circulation zone.

According to the magnitude of resonant Hamiltonian, there are two special cases. The first case is that, when ${\cal H}$ is equal to that of the resonant center, the magnitude of $S$ becomes zero, meaning that the small body is initially placed at the resonant center. As discussed in the introduction, such a special case with $S=0$ has been adopted as an assumption in formulating secular models \citep{kozai1985secular, yoshikawa1989survey, nesvorny2002perturbative, wan2007exploration, saillenfest2017study, li2021apsidal, Pons2022Secular}. The second case is that, when ${\cal H}$ is equal to that of the saddle point (or the dynamical separatrix), $S$ is equal to the area bounded by the dynamical separatrices and, in this case, the assumption that the two degrees of freedom are separable in frequencies is failed, leading to the fact that $S$ is no longer an adiabatic invariant \citep{wisdom1985perturbative, neishtadt1987change, neishtadt2004wisdom, tennyson1986change}. In the following, we denote the curve on which the Hamiltonian ${\cal H}$ is equal to that of the saddle point ${\cal H}_U$ as the critical curve. From the phase portrait shown in Fig. \ref{Fig4}, we can see that the region with ${\cal H} > {\cal H}_U$ corresponds to libration and the region with ${\cal H} < {\cal H}_U$ corresponds to circulation. The phase-space regions around critical curve are called zones of uncertainty and crossing of uncertainty zones is a generator of chaos \citep{wisdom1985perturbative}.

In Fig. \ref{Fig5}, the geometrical definition of $S$ is illustrated for resonances of 1:n type. In practice, we take 1:2 resonance with Neptune as an example. In the phase portraits of MMR, the dynamical separatrices corresponding to level curves of ${\cal H} = {\cal H}_U^{(1)}$ (inner separatrix) and ${\cal H} = {\cal H}_U^{(2)}$ (outer separatrix) are shown by red dashed lines. In Fig. \ref{Fig5}, the regions enclosed by level curves of given resonant Hamiltonian are marked by shaded zones, and the area of these zones corresponds to the adiabatic invariant $S$. According to the magnitude of resonant Hamiltonian, there are three cases:
\begin{description}
\item[Case I: Inside the leading or trailing island] When the resonant Hamiltonian ${\cal H}$ is greater than ${\cal H}_U^{(1)}$, the particle is located inside the asymmetric libration islands (the leading island is on the left and the trailing island is on the right). In this case, there are two isolines of ${\cal H}$, one is inside the leading island and the other one is inside the trailing island. Inside the asymmetric libration islands, the trajectories are of tadpole type. Please refer to panels (a) and (b) of Fig. \ref{Fig5}.
\item[Case II: Inside the symmetric islands] When the resonant Hamiltonian ${\cal H}$ is smaller than ${\cal H}_U^{(1)}$ but greater than ${\cal H}_U^{(2)}$, the particle is located inside the symmetric libration island bounded by the inner and outer separatrices. In this case, both the asymmetric islands are bounded by the isoline of resonant Hamiltonian and the trajectories in the symmetric libration island are of horseshoe type. Please see panel (c) of Fig. \ref{Fig5}.
\item[Case III: Outside the resonant regions] When the resonant Hamiltonian ${\cal H}$ is smaller than ${\cal H}_U^{(2)}$, test particles are located outside the resonant zones. Please see panel (d) of Fig. \ref{Fig5}.
\end{description}

In the long-term evolution of TNOs under the considered two-degree-of-freedom dynamical model, both the resonant Hamiltonian ${\cal H}$ and the adiabatic invariant $S$ remain unchanged. Thus, it is not difficult to conclude that, when the parameters $I_{\max}$ and ${\cal H}$ are provided, the motion of TNOs takes place on the isolines of the adiabatic invariant $S$ in the phase space spanned by $(\omega, \tilde{e})$. Based on this fact, it is possible to produce phase portraits for secular evolutions by plotting level curves of the adiabatic invariant $S$ with given ${\cal H}$ \citep{wisdom1985perturbative, henrard1989motion, henrard1990semi, neishtadt2004wisdom} or by plotting level curves of Hamiltonian ${\cal H}$ with given $S$ \citep{morbidelli2002modern, saillenfest2016long, saillenfest2020long}. By analyzing phase portraits, it is possible to investigate long-term dynamics for TNOs inside or around MMRs. This is the semi-analytical secular model formulated in this study.

\begin{figure*}
\centering
\includegraphics[width=0.45\textwidth]{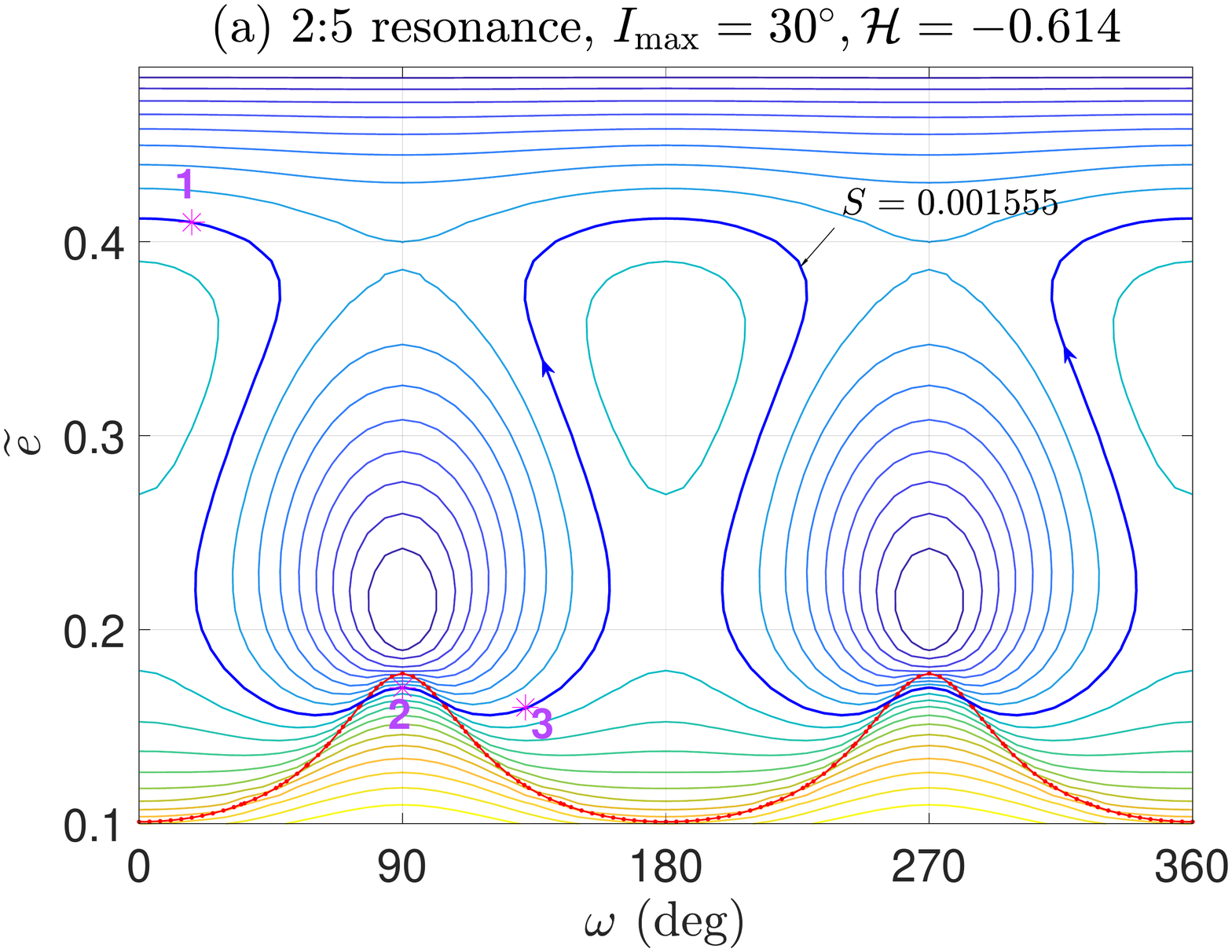}
\includegraphics[width=0.45\textwidth]{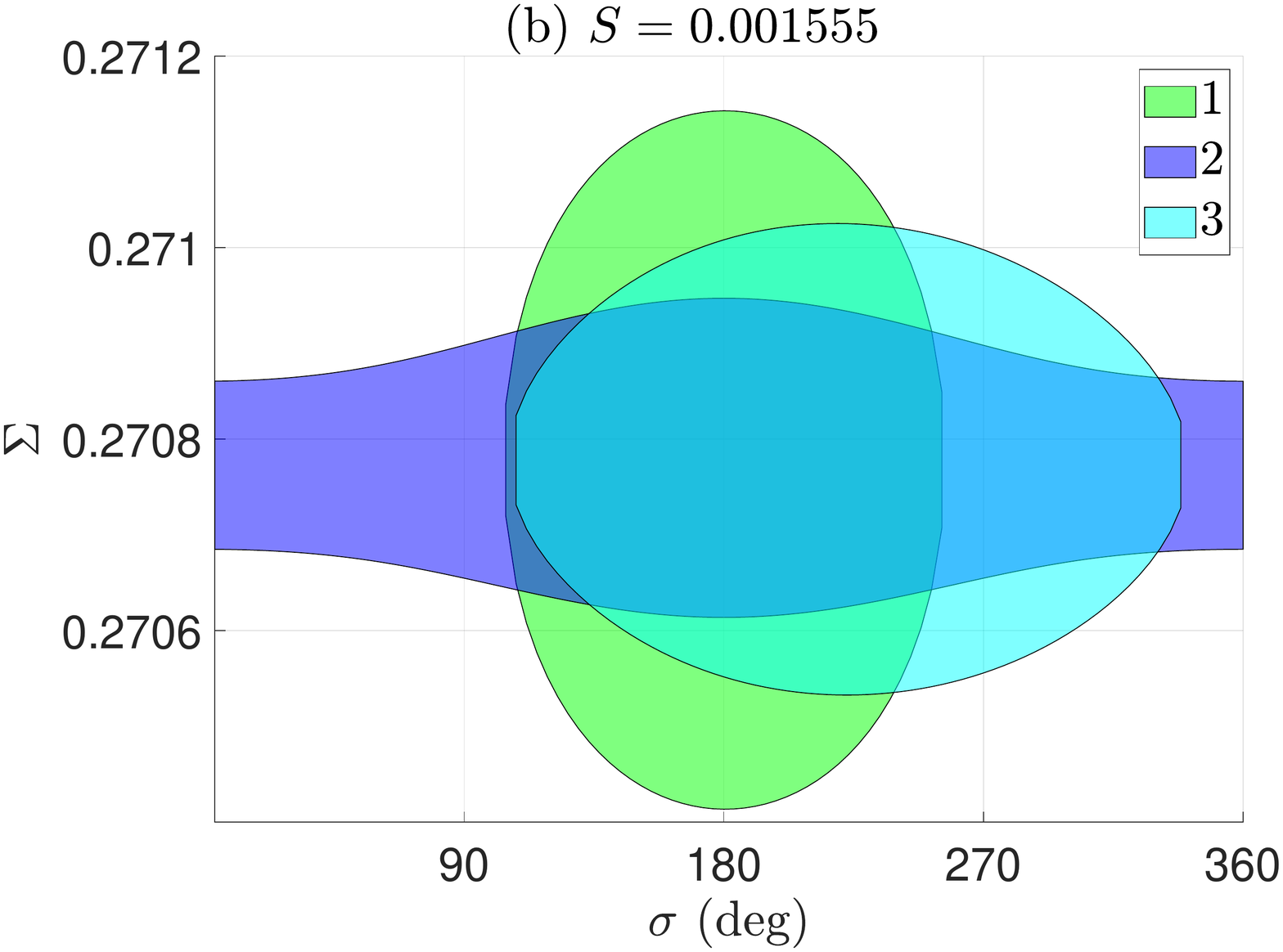}
\caption{Level curves of the adiabatic invariant, $S$, plotted in the space $(\omega, \widetilde{e})$ (\emph{left panel}), and three representative examples with $S=0.001555$, plotted in the $(\sigma, \Sigma)$ space (\emph{right panel}). In the \emph{left panel}, the level curve of $S=0.001555$ is shown by a blue line and the critical curve specified by ${\cal H}_U = -0.614$ is shown by a red line. In particular, the region above the uncertainty curve corresponds to libration while the region below the critical curve corresponds to circulation. In the \emph{right panel}, the magnitude of $S$ stands for the area of the shaded region bounded by the isolines of resonant Hamiltonian. Following the blue line shown in the \emph{left panel}, particles could switch between libration and circulation regions periodically.}
\label{Fig6}
\end{figure*}

To clearly illustrate our semi-analytical model, we take 2:5 resonance with Neptune as an example to plot level curves of the adiabatic invariant $S$ in the space $(\omega, \tilde{e})$ by taking $I_{\max} = 30^{\circ}$ and ${\cal H} = -0.614$. The results are shown in the left panel of Fig. \ref{Fig6}. The red dotted line represent the critical curve specified by ${\cal H}_U = -0.614$. The region below the critical curve corresponds to circulation (because in this region the Hamiltonian is smaller than that of the separatrix) and the region above the critical curve corresponds to libration (because in this region the Hamiltonian is greater than that of the separatrix). Observing from the phase portrait shown in the left panel of Fig. \ref{Fig6}, we can see that the level curves of $S$ in the libration and circulation regions are continuous, as desired. In the long-term evolution, the test particle moves along a certain level curve shown in the phase portrait. As an example, the level curve of $S=0.001555$ is explicitly marked by a blue line in the phase portrait and three typical points on the level curve are picked out and marked by pink stars. These points are denoted by numbers 1, 2 and 3. It is observed that points 1 and 3 are located inside the libration regions (i.e., above the critical curve) and point 2 is located inside the circulation region (i.e., below the critical curve). In the right panel of Fig. \ref{Fig6}, geometrical definitions of the adiabatic invariant $S$ corresponding to these three points are illustrated in the space $(\sigma, \Sigma)$. Moving from point 1 to point 2, the particle transits from libration region to circulation region and then return to libration region when moving from point 2 to point 3. Theoretically speaking and according to this model, a particle moving along this blue line will switch between libration and circulation periodically.

From the right panel of Fig. \ref{Fig6}, it is observed that, in the long-term evolution, the resonant angle $\sigma$ presents a significant variation in terms of both the resonant center and libration amplitude. This indicates that some artificial assumptions for the variation of $\sigma$ in previous works (e.g., zero or constant amplitude of $\sigma$) may lead to unreliable results. This problem was noted by \citet{brasil2014dynamical} and \citet{saillenfest2016long}.

From the phase portrait presented in the left panel of Fig. \ref{Fig6}, we could find some interesting dynamical structures: (a) both the arguments $\sigma$ and $\omega$ circulate in the regions below the critical curve, (b) ZLK resonances can be found in the region above the critical curve, (c) the ZLK islands centered at $\omega=90^{\circ}$ or $\omega=270^{\circ}$ are larger than the one centered at $\omega = 0$ or $\omega = \pi$, (d) the ZLK resonance centered at $\omega = 0$ or $\omega = \pi$ occupies a higher-eccentricity space, and (e) the ZLK resonance disappears in the region with eccentricities greater than $\sim$0.41.

\begin{figure*}
\centering
\includegraphics[width=0.45\textwidth]{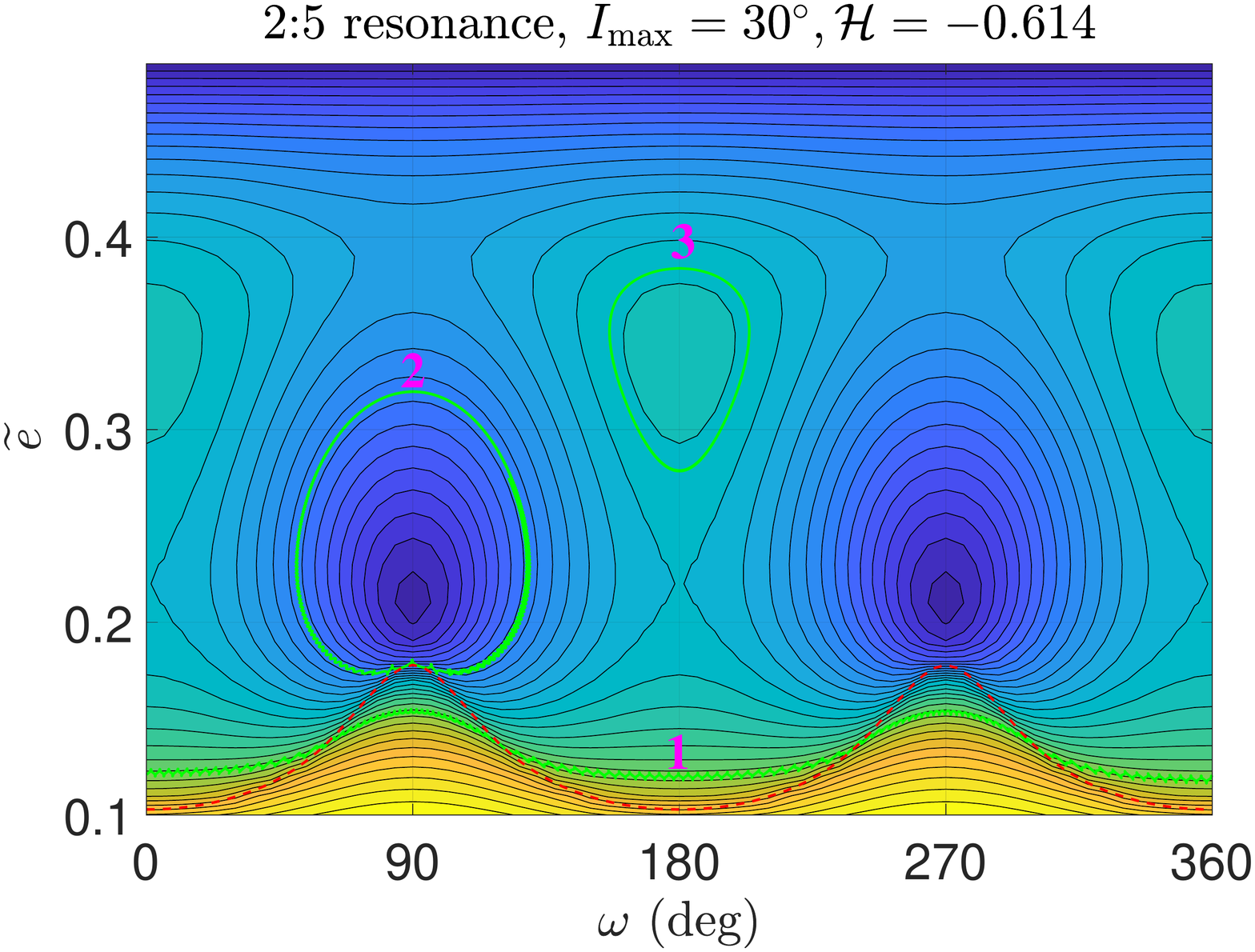}
\includegraphics[width=0.45\textwidth]{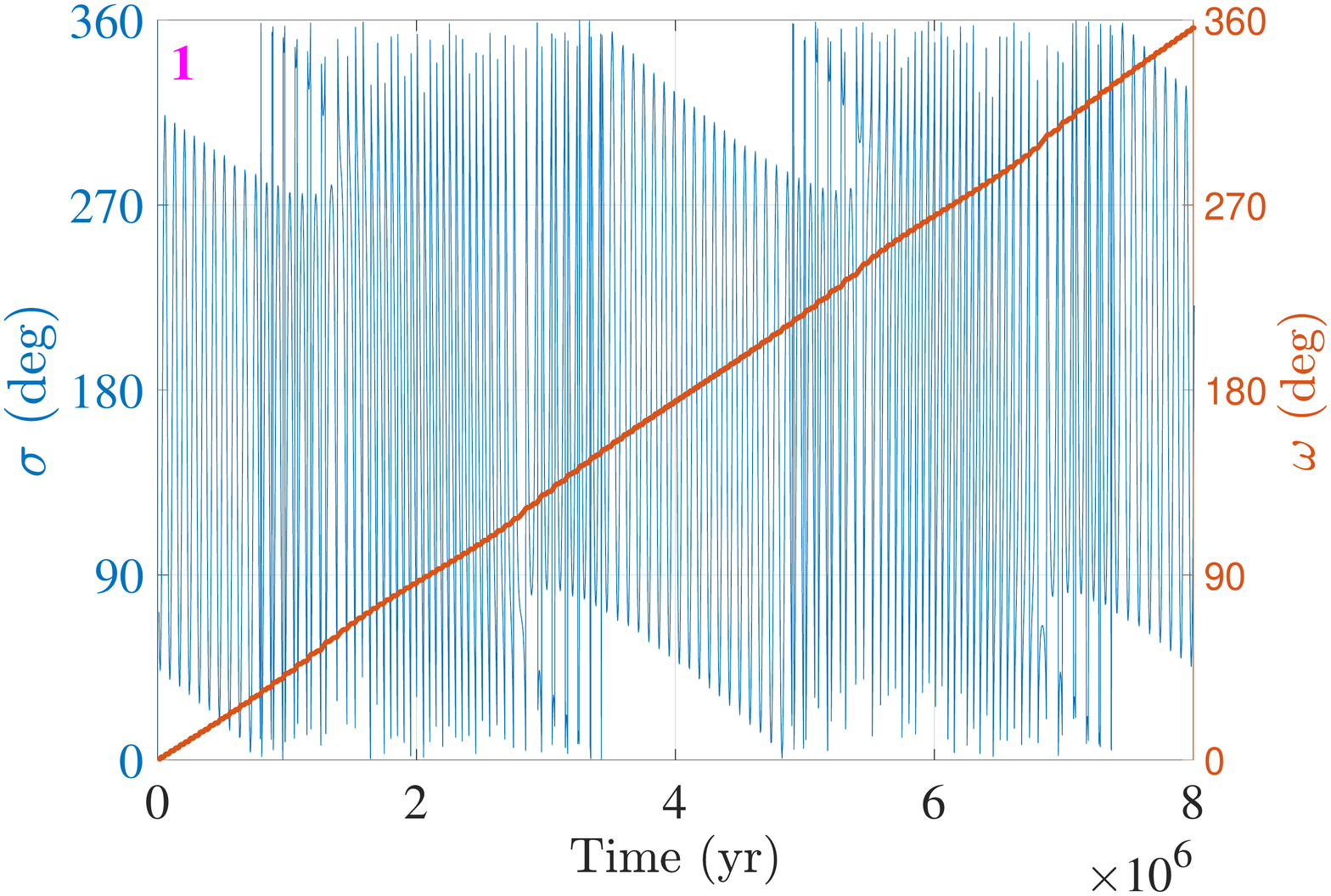}\\
\includegraphics[width=0.45\textwidth]{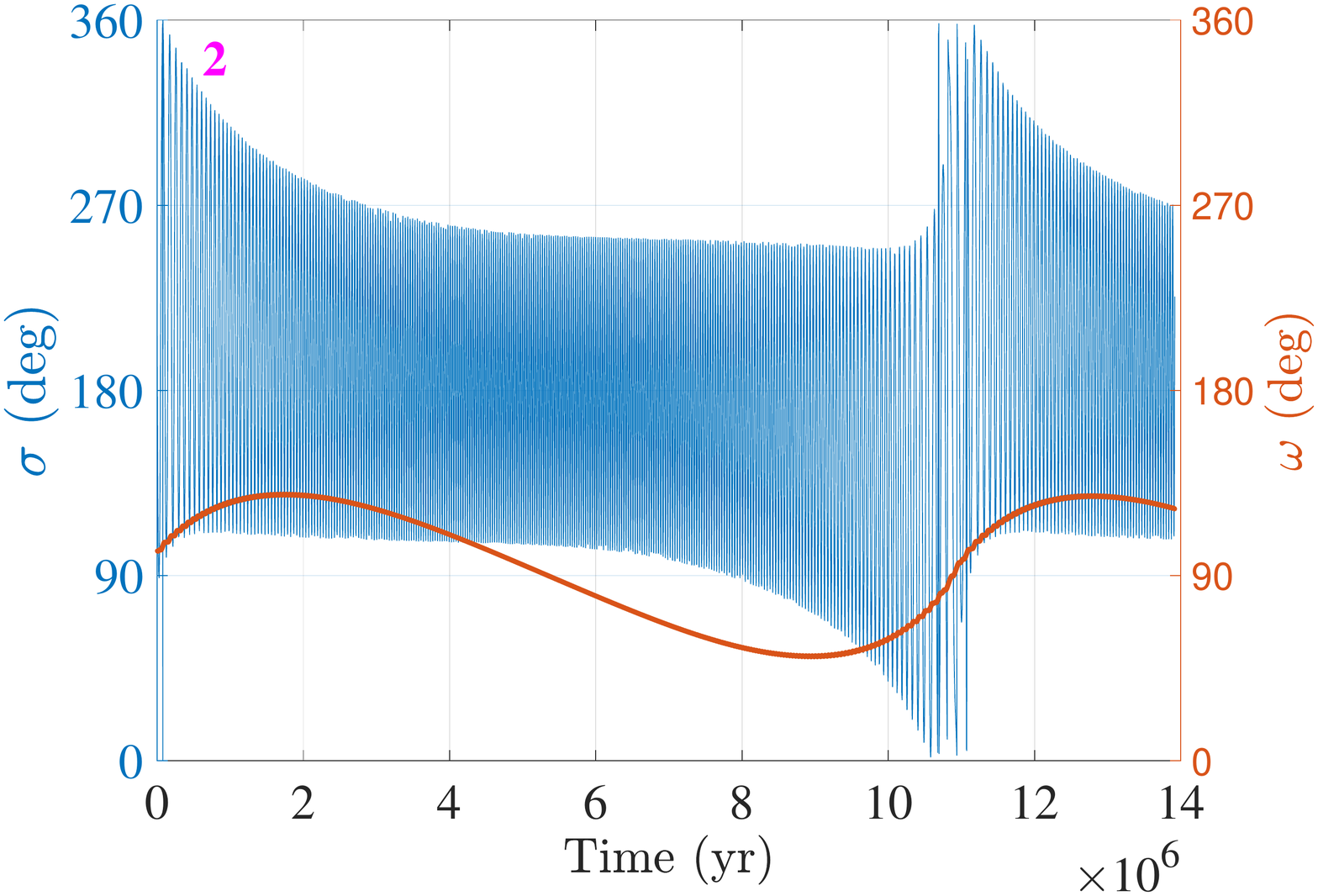}
\includegraphics[width=0.45\textwidth]{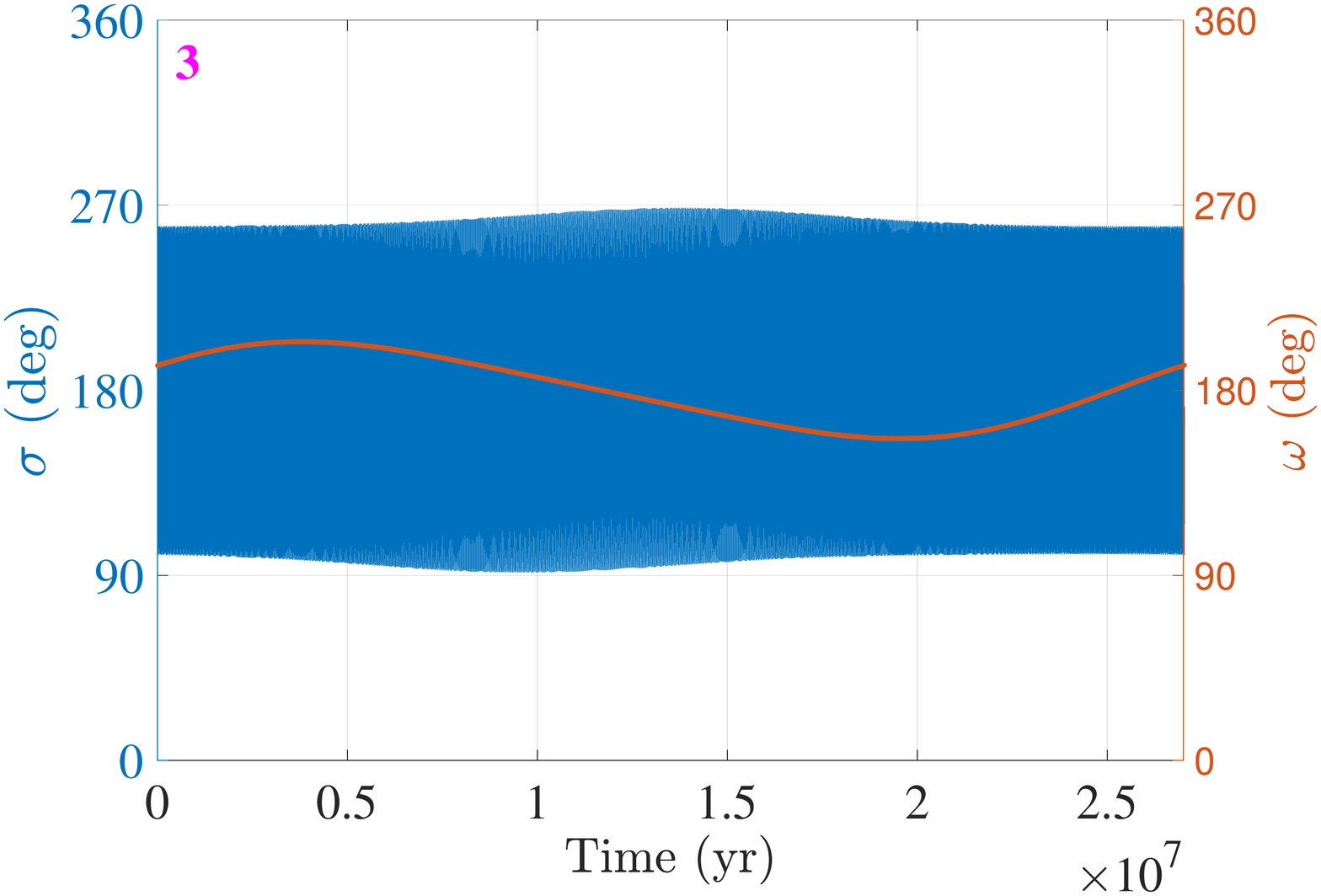}
\caption{Level curves of the adiabatic invariant $S$ (i.e., phase portrait) in the secular evolution for TNOs inside 2:5 resonance with Neptune (\emph{the upper-left panel}) and time histories of $\sigma$ and $\omega$ of three representative trajectories numerically propagated in the semi-secular model (\emph{the remaining three panels}). In practical simulations, the maximum inclination is fixed at $I_{\max} = 30^{\circ}$ and the resonant Hamiltonian is assumed at ${\cal H} = -0.614$. In the phase portrait (see the upper-left panel), three numerically integrated trajectories are shown by green lines and denoted by numbers 1, 2 and 3, and the critical curve specified by ${\cal H}_U = -0.614$ is marked by a red line. It is observed that the numerically propagated trajectories follow closely along the level curves in the phase portrait.}
\label{Fig7}
\end{figure*}

When the particle is crossing the ``zone of uncertainty", the adiabatic invariant approximation is invalidated and chaos may occur \citep{wisdom1985perturbative}. However, for the current dynamical model, the zone of uncertainty is very narrow, meaning that the change of adiabatic invariant $S$ is negligible and the value of $S$ is still predictable for each possible transition. Thus, whether the particles are inside or around a certain MMR, the level curves of the adiabatic invariant $S$ are available to predict secular behaviors. About this problem, please see \citet{saillenfest2016long} for a detailed discussion.

\begin{figure*}
\centering
\includegraphics[width=0.45\textwidth]{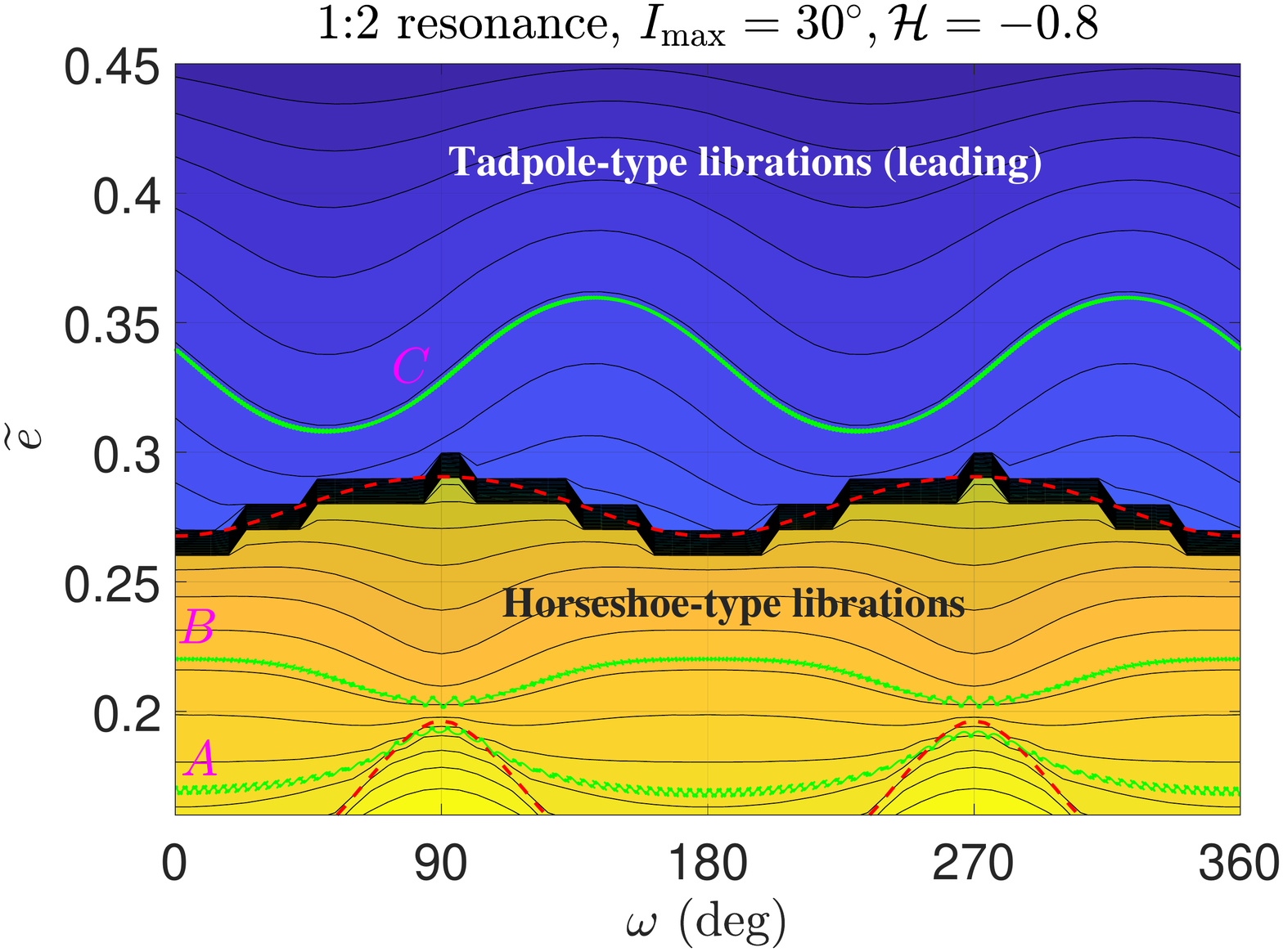}
\includegraphics[width=0.45\textwidth]{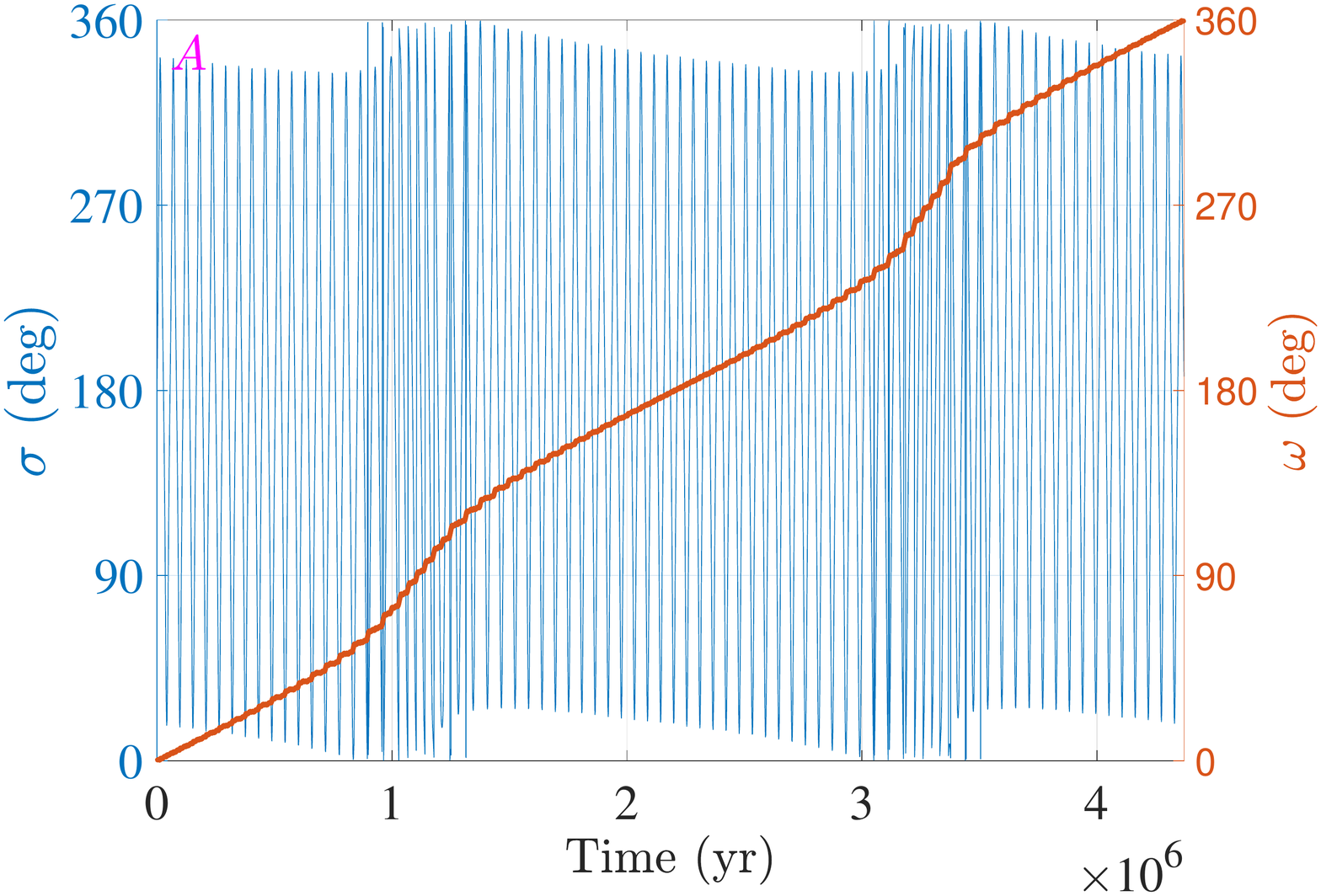}\\
\includegraphics[width=0.45\textwidth]{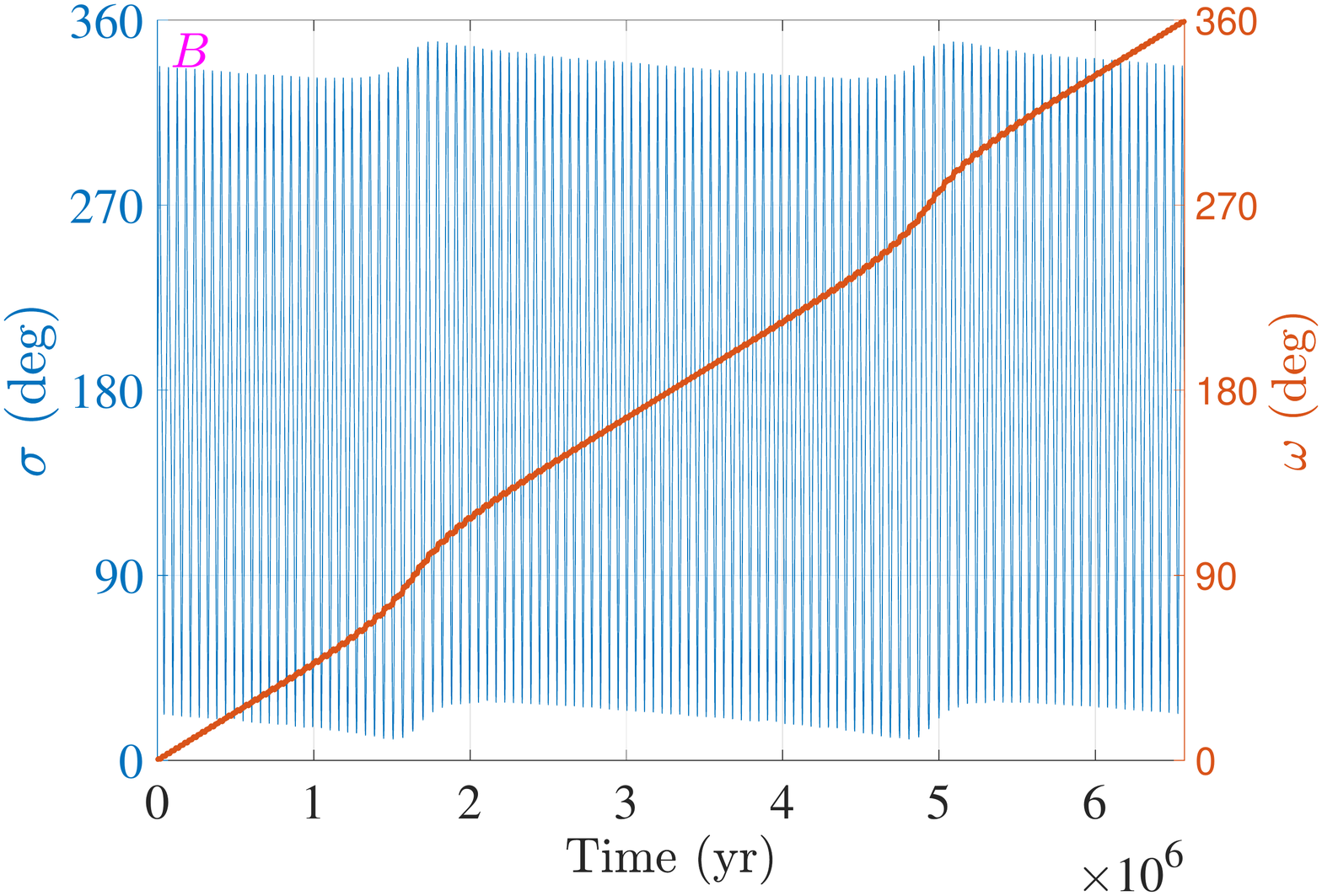}
\includegraphics[width=0.45\textwidth]{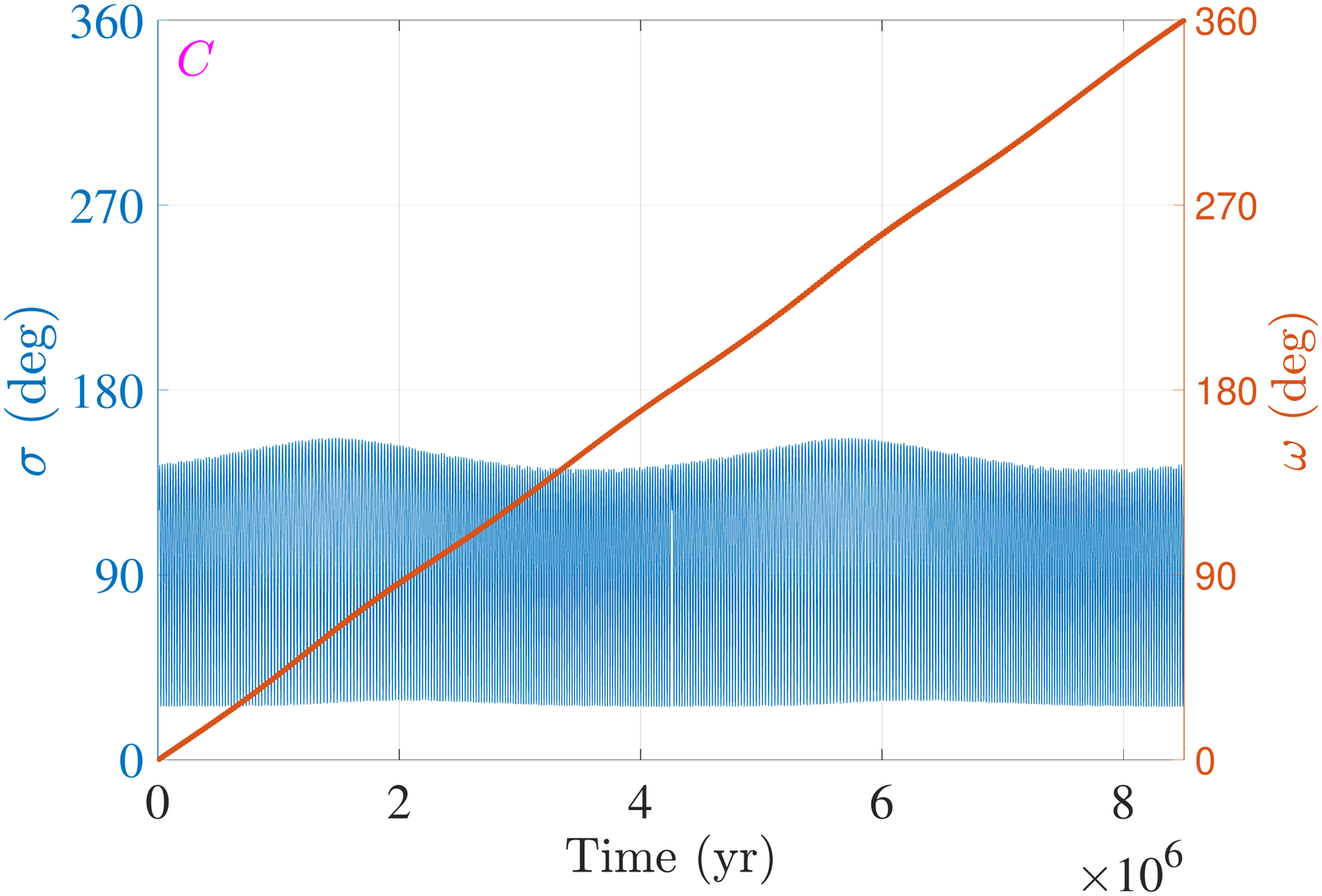}
\caption{Level curves of the adiabatic invariant $S$ (i.e., phase portrait) in the secular evolution for minor bodies inside 1:2 resonance with Neptune (\emph{the upper-left panel}) and time histories of $\sigma$ and $\omega$ of three representative trajectories numerically propagated in the semi-secular model (\emph{the remaining three panels}). In the tadpole-type region, the adiabatic invariant $S$ is measured at the leading island. In the phase portrait (see the upper-left panel), three representative trajectories, denoted by letters A, B and C, are shown by green lines and the critical curves are marked by red dashed lines.}
\label{Fig8}
\end{figure*}

\begin{figure*}
\centering
\includegraphics[width=0.45\textwidth]{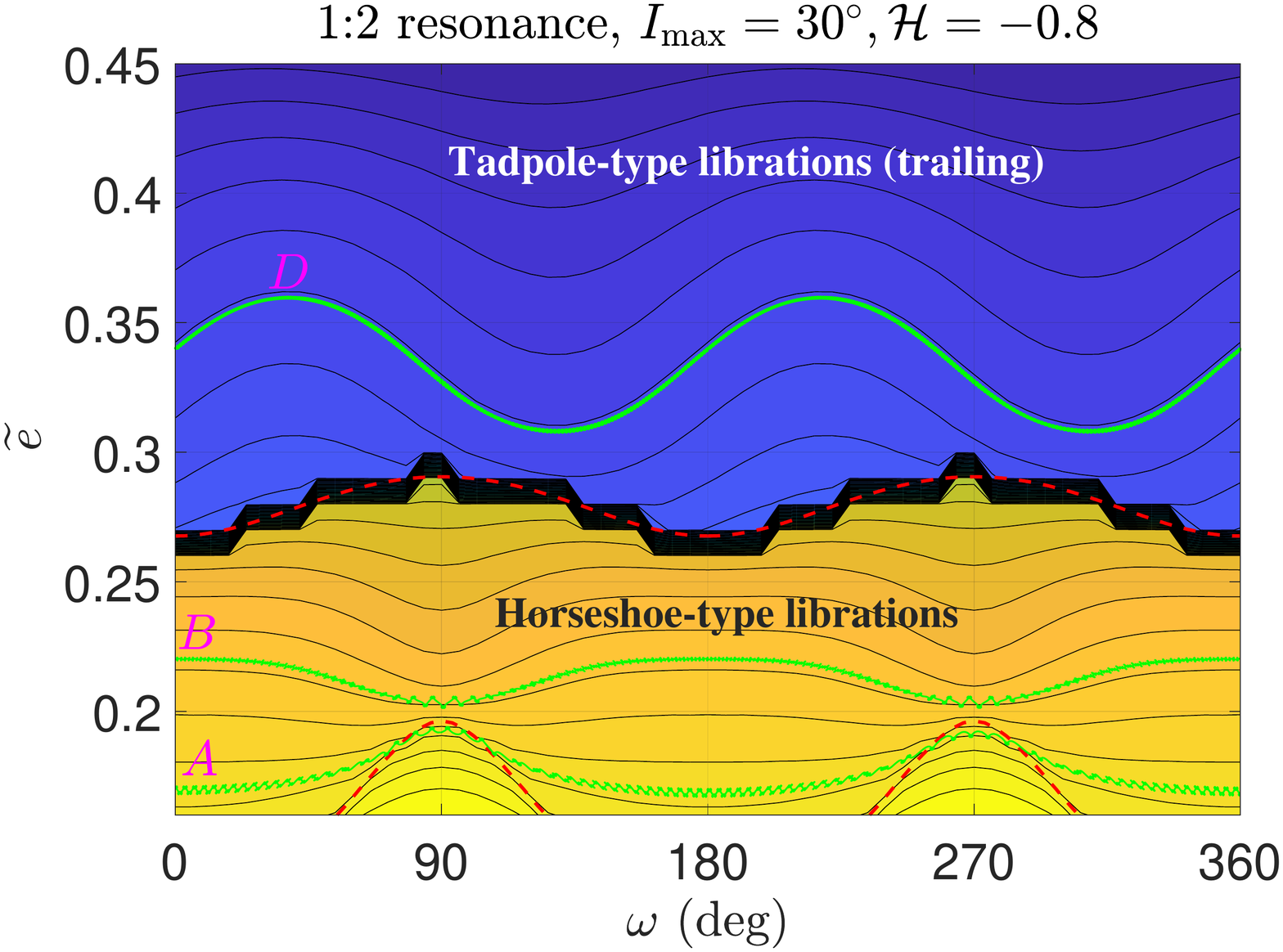}
\includegraphics[width=0.45\textwidth]{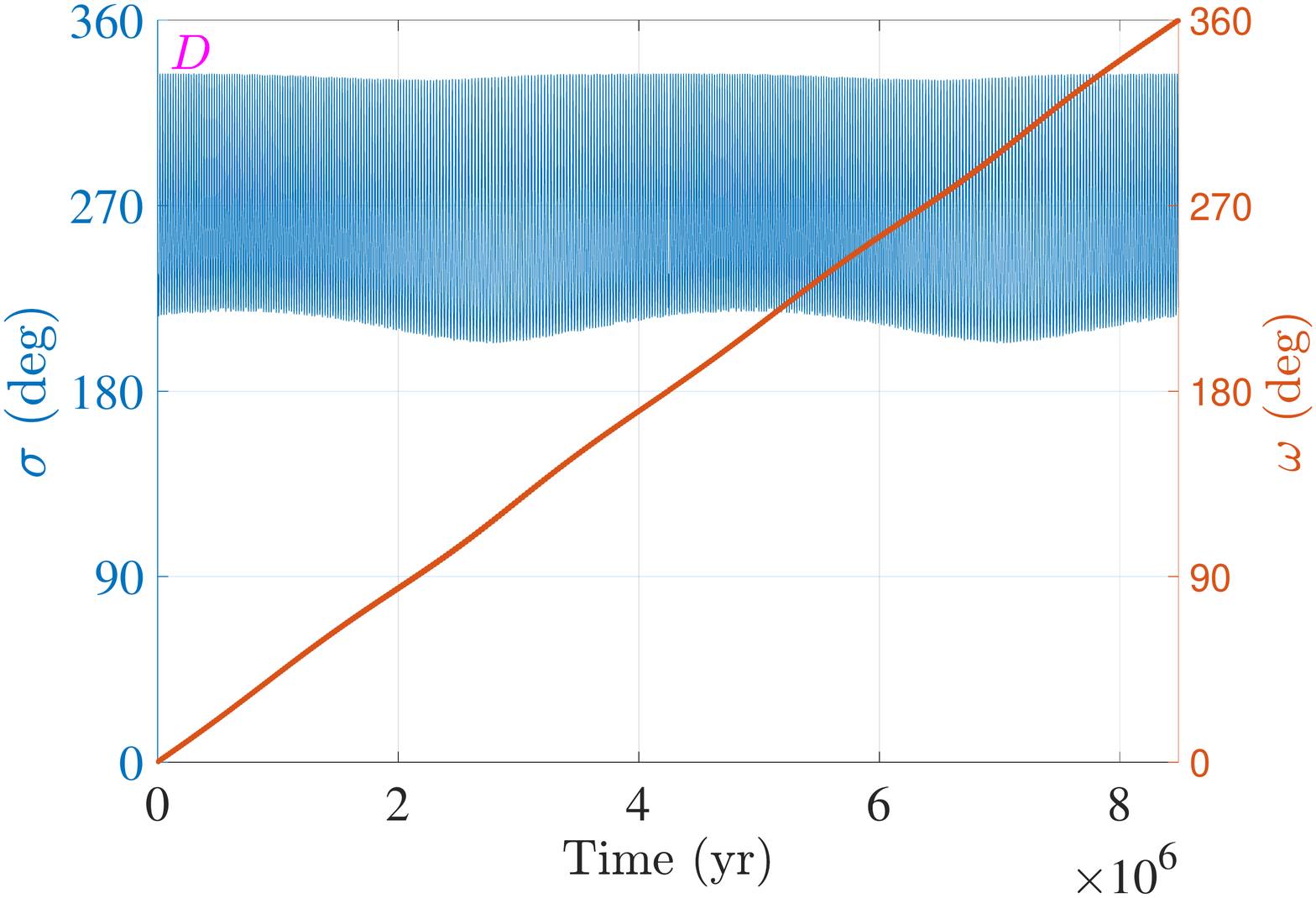}
\caption{Level curves of the adiabatic invariant $S$ (i.e., phase portrait) in the secular evolution for minor bodies inside 1:2 resonance with Neptune (\emph{the left panel}) and time histories of $\sigma$ and $\omega$ of the trajectory D numerically propagated in the semi-secular model (\emph{the right panel}). In the phase portrait (see the left panel), three representative trajectories, denoted by letters A, B and D, are shown by green lines and the critical curves are marked by red dashed lines. It is noted that the time histories of $\sigma$ and $\omega$ of trajectories A and B have been shown in Fig.~\ref{Fig8}. In the tadpole-type region, the adiabatic invariant $S$ is measured at the trailing island.}
\label{Fig9}
\end{figure*}

\subsection{Validation of the semi-analytical developments}
\label{Sect4-2}

Before applications to real TNOs, we need to validate the semi-analytical model by comparing the trajectories of particles predicted by phase portraits with the ones numerically propagated under the semi-secular model. For convenience, the resonances of non-1:n type and the ones of 1:n type are discussed separately.

For the resonances of non-1:n type, we take 2:5 resonance with Neptune as an example. Following the discussion presented in the previous subsection, we produce the phase portrait by plotting the level curves of $S$ with the parameters $I_{\max} = 30^{\circ}$ and ${\cal H} = -0.614$, as shown in the upper-left panel of Fig. \ref{Fig7}. In the phase portrait, the critical curve is marked in red line. The region below the critical curve corresponds to circulation and the region above the critical curve corresponds to libration. In addition, with the same $I_{\max}$ and ${\cal H}$, three trajectories are numerically propagated under the semi-secular model and the resulting numerical trajectories are denoted by numbers 1, 2 and 3, as shown by green lines. To show the detailed behaviors of evolution, the time histories of $\sigma$ and $\omega$ for these numerically propagated trajectories are reported in the remaining three panels of Fig. \ref{Fig7}.

A good agreement can be observed between the level curves shown in the phase portrait and the trajectories numerically propagated under the semi-secular model, as shown in the first panel of Fig. \ref{Fig7}. This shows that the adiabatic invariance approximation works very well for the problem at hand.

The remaining three panels of Fig. \ref{Fig7} indicate that (a) following trajectory 1 the particle switches between libration and circulation in terms of $\sigma$ and it is located outside the ZLK resonance, (b) following trajectory 2 the particle also switches between libration and circulation in terms of $\sigma$ but it is located inside the ZLK resonance with $\omega$ librating around $90^{\circ}$ and (c) following trajectory 3 the particle is located inside both mean motion resonance and the ZLK resonance (both $\sigma$ and $\omega$ are librating around $180^{\circ}$).

As for the resonances of 1:n type, we take 1:2 resonance with Neptune as an example. To produce the phase portrait, we take the following parameters: $I_{\max} = 30^{\circ}$ and ${\cal H} = -0.8$. The level curves of the adiabatic invariant $S$ shown in the space $(\omega, \tilde{e})$ are reported in the upper--left panels of Figs \ref{Fig8} and \ref{Fig9}. In the phase portrait, the critical curves are marked by red dashed lines. The critical curves divide the entire phase space into three subregions: the first one with circulating trajectories, the second one with horseshoe-type trajectories and the last one with tadpole-type trajectories. In the phase portrait shown in Fig. \ref{Fig8} the adiabatic invariant $S$ is evaluated inside the leading island and it is evaluated inside the trailing island for the phase portrait shown in Fig. \ref{Fig9}.

Under the semi-secular model, four trajectories are numerically propagated and they are denoted by letters A, B, C and D, respectively. Their time histories of $\sigma$ and $\omega$ are reported in the last three panels of Fig. \ref{Fig8} and in the right panel of Fig. \ref{Fig9}.

Also, an excellent agreement is observed between level curves in the phase portraits and the corresponding numerically propagated trajectories (see the first panels of Figs \ref{Fig8} and \ref{Fig9}). It further indicates that the semi-analytical secular model formulated in this study is applicable for predicting long-term evolution for TNOs inside or around MMRs.

Observing Figs \ref{Fig8} and \ref{Fig9}, we can see that (a) following trajectory A the particle transits between the libration region with horseshoe-type trajectories and the circulation region, (b) following trajectory B the particle is inside the libration region with horseshoe-type trajectories, (c) following trajectory C the particle is inside the leading island and (d) following trajectory D the particle is inside the trailing island. The numerical behaviors are in quite good agreement with the ones predicted by phase portraits.

\begin{table*}
\small
\centering
\caption{Averaged elements (including semimajor axis $a$, eccentricity $e$, inclination $I$, longitude of ascending node $\Omega$, argument of perihelion $\omega$ and the resonant argument $\sigma$), the maximum inclination $I_{\max}$ and resonant Hamiltonian $\cal H$ for the representative TNOs inside MMRs with Neptune considered in this study. To evaluate the averaged elements, the initial epoch is taken on December 17th, 2020.}
\begin{tabular*}{\hsize}{@{}@{\extracolsep{\fill}}lcccccccccc@{}}
\hline
TNO & $k_p$:$k$ & $a$ (au) & $e$ & $I$ (deg) & $\Omega$ (deg) & $\omega$ (deg) & $\sigma$ (deg) & $\cal H$ & $I_{\max}$ (deg) & Figure\\
\hline\hline
2018 VO$_{137}$ & 2:5 &55.43896&0.187185&38.99327&42.69996&135.01319& 114.97992&$-$0.584742&40.20489& Fig. \ref{Fig10}\\
\hline
2005 SD$_{278}$ & 2:5 &55.58254&0.28142&17.89926&152.53972& 218.34871&188.25311&$-$0.62790&23.91183& Fig. \ref{Fig11}(a)\\
2015 PD$_{312}$ & 2:5 &55.48197&0.374209&23.01589&154.78276& 226.08094&105.88611&$-$0.60659&31.34801& Fig. \ref{Fig11}(b)\\
\hline
Pluto           & 2:3 &39.56190&0.24941&17.14126&110.25738& 113.79771&235.18742&$-$0.92683&22.18451& Fig. \ref{Fig12}(a)\\
2004 HA$_{79}$  & 2:3 &39.37833&0.24501&22.69980&203.20489& 262.65667&161.56184&$-$0.84521&26.56732& Fig. \ref{Fig12}(b)\\
\hline
1996 TR$_{66}$  & 1:2 &47.89136&0.398146&12.42382&343.04016& 309.39781&58.14003&$-$0.68111&26.28515& Fig. \ref{Fig13}(a)\\
2014 SR$_{373}$ &1:3  &62.55550&0.38324&35.58437&165.74151& 214.52424&80.01628&$-$0.49380&41.29682& Fig. \ref{Fig13}(b)\\
\hline
\end{tabular*}
\label{Tab2}
\end{table*}

\begin{figure*}
\centering
\includegraphics[width=0.45\textwidth]{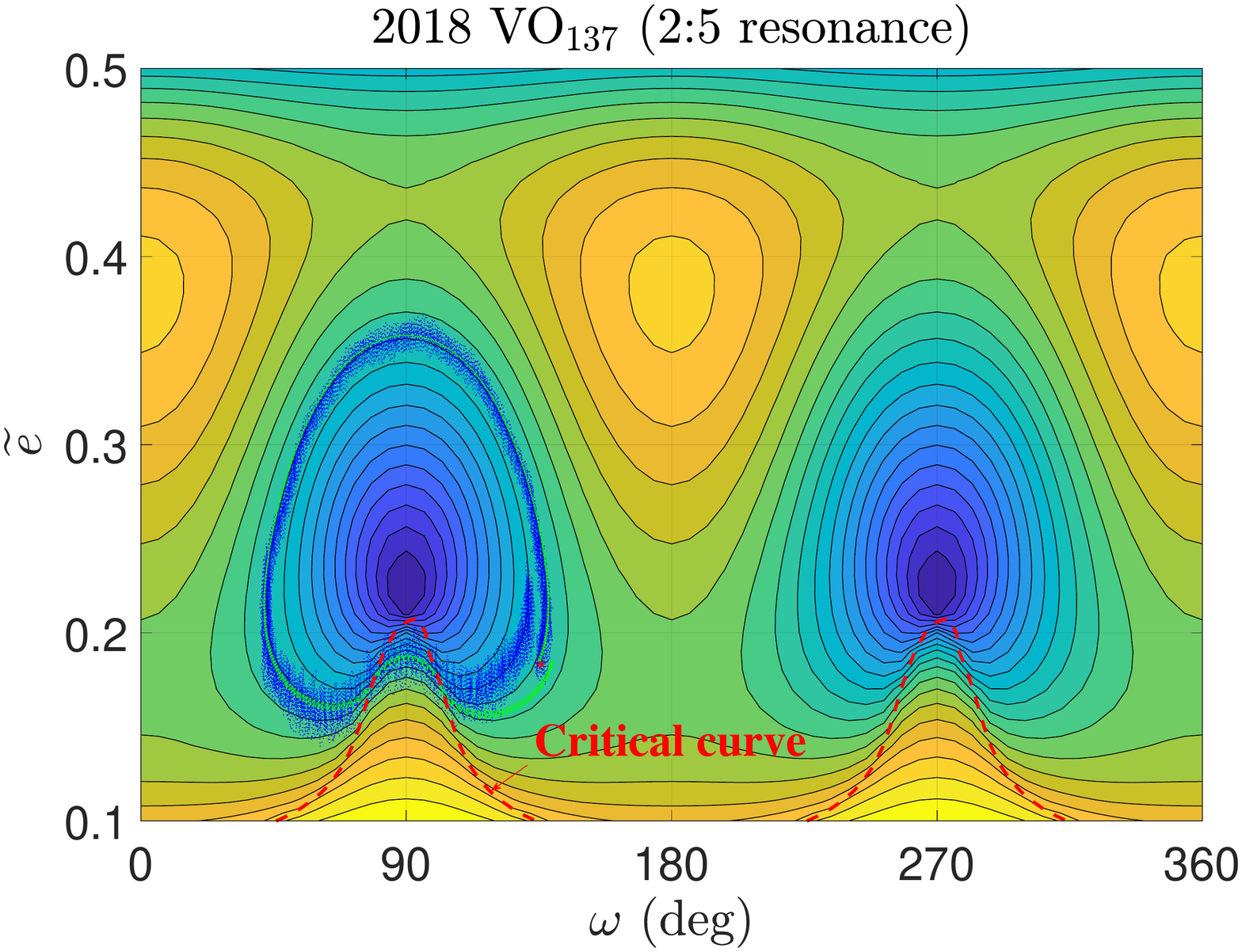}
\includegraphics[width=0.45\textwidth]{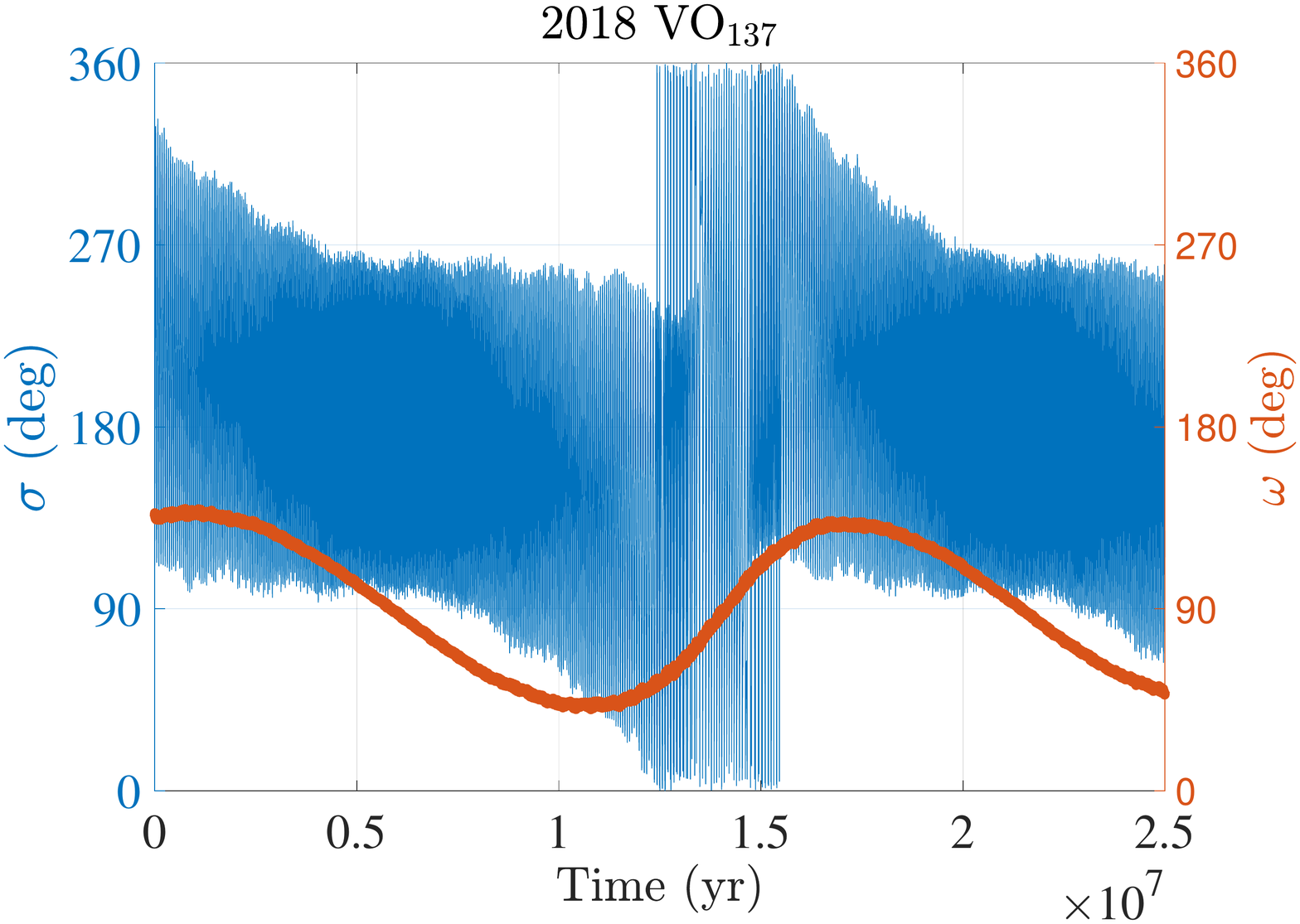}
\caption{Level curves of the adiabatic invariant $S$ (\emph{left panel}) and time histories of the angles $\sigma$ and $\omega$ for the trajectory propagated under the full $N$-body model (\emph{right panel}) about 2018 VO$_{137}$, which is currently inside 2:5 resonance with Neptune. The parameters about this TNO are provided in Table \ref{Tab2}. In the \emph{left panel}, the trajectory propagated in the full $N$-body model is shown in blue line and the one propagated in the semi-secular model is shown in green line. The critical curve is marked by red dots (the region above the curve corresponds to libration and the one below this curve corresponds to circulation). Evidently, the trajectory of 2018 VO$_{137}$ intersects the critical curve twice in a ZLK period, showing that in the long-term evolution the resonant angle $\sigma$ switches between libration and circulation periodically. In the considered time interval, 2018 VO$_{137}$ is currently located inside the ZLK island centered at $\omega = 90^{\circ}$.}
\label{Fig10}
\end{figure*}

\begin{figure*}
\centering
\includegraphics[width=0.45\textwidth]{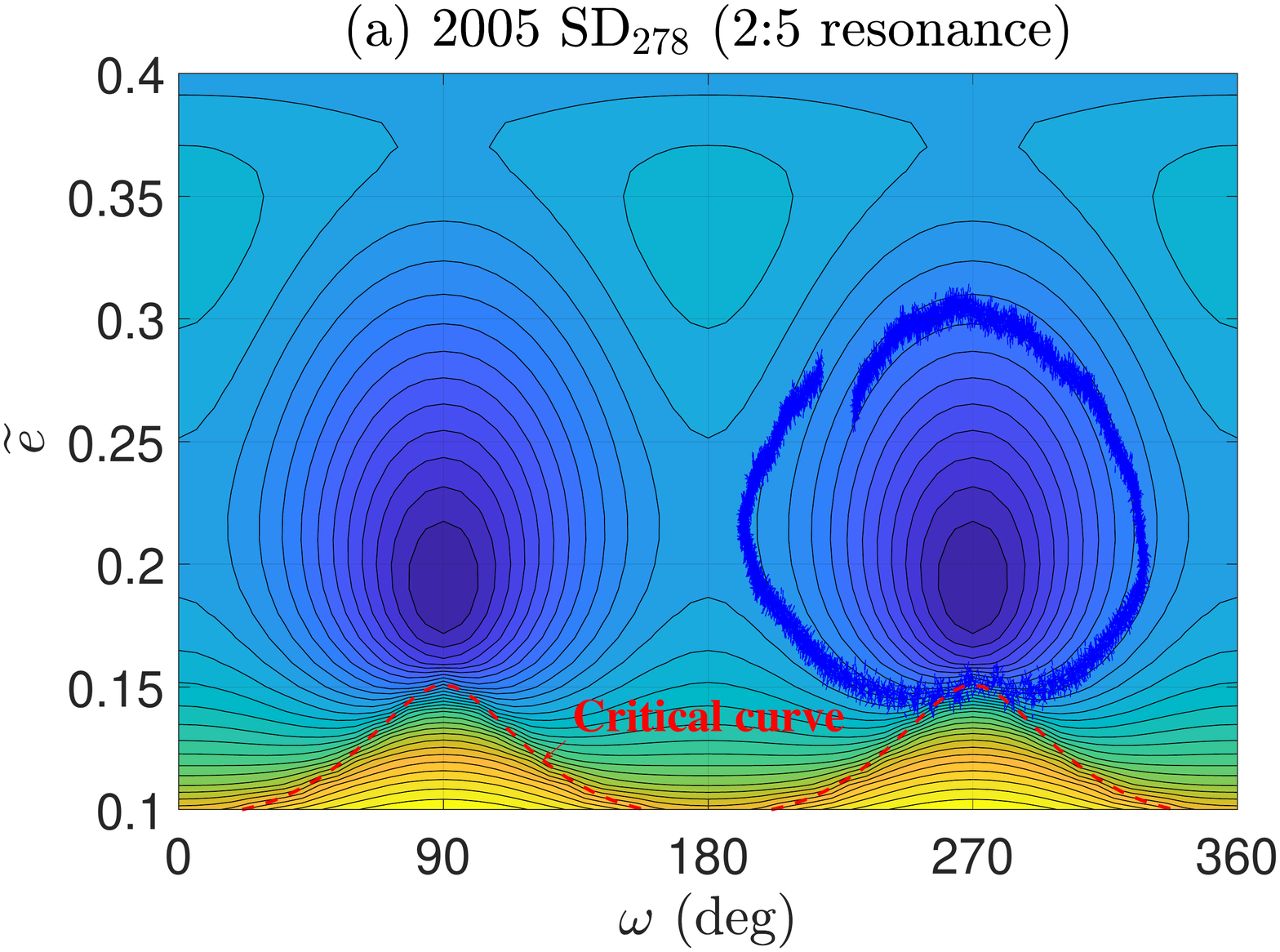}
\includegraphics[width=0.45\textwidth]{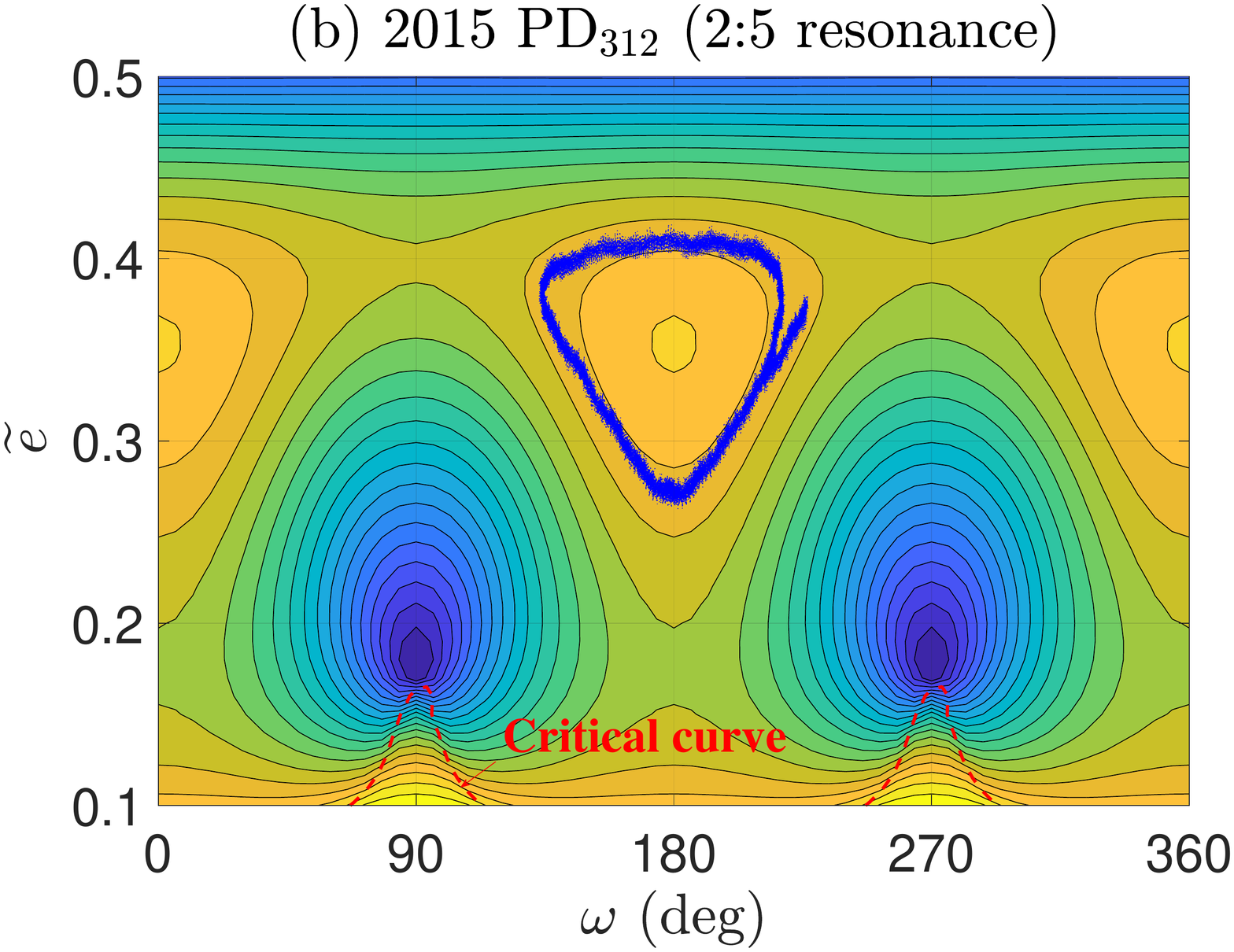}
\caption{Similar to the upper-left panel of Fig. \ref{Fig10} but for another two representative TNOs (2005 SD$_{278}$ and 2015 PD$_{312}$) inside 2:5 resonance with Neptune. During the considered period of time, these two TNOs are also located inside the ZLK resonance.}
\label{Fig11}
\end{figure*}

\section{Applications to real trans-Neptunian objects}
\label{Sect5}

In this section, we apply the semi-analytical secular model formulated in Section \ref{Sect4} to real TNOs inside MMRs with Neptune\footnote{https://minorplanetcenter.net, retrieved 9 April 2021}. In practical applications, we take representative members from several groups of resonant TNOs as examples. It should be mentioned that the secular model developed in this work is also applicable for other resonant objects. According to the average defined by Eq. (\ref{Eq9}), averaged elements of TNOs are required in the semi-secular dynamical model. To produce averaged elements of TNOs, we numerically integrate the equations of motion under the full $N$-body model over $k$ times orbital periods of Neptune to obtain the osculating elements, and then we identify the averaged elements of TNOs by numerically averaging the time series of osculating elements over the duration of integration. Table \ref{Tab2} provides the averaged elements, the maximum inclination $I_{\max}$ and resonant Hamiltonian ${\cal H}$ for the TNOs chosen in this study. It should be noted that, in the process of evaluating the averaged elements, the initial epoch is taken on December 17th, 2020.

In Fig. \ref{Fig10}, we take into account the secular dynamics of 2018 VO$_{137}$, which is currently inside 2:5 resonance. For this TNO, its maximum inclination is $I_{\max}= 40.2^{\circ}$ and the resonant Hamiltonian is ${\cal H} = -0.58$ (see Table \ref{Tab2}). With the given parameters $I_{\max}$ and ${\cal H}$, the associated phase portrait is presented in the left panel of Fig. \ref{Fig10}. For convenience, the critical curve is marked by a red dashed line. Recall that the critical curve is defined as the curve on which the resonant Hamiltonian is equal to that of the dynamical separatrix. As stated before, the phase-space regions below the critical curve are of circulation and the ones above the critical curve are of libration. The trajectories are numerically propagated under both the semi-secular model and under the full $N$-body model. The trajectory propagated under the semi-secular model is given by green line and the one produced under the full $N$-body model is shown in blue line (see the left panel of Fig. \ref{Fig10} for details). In the right panel of Fig. \ref{Fig10}, the time histories of $\sigma$ and $\omega$ are presented for the trajectory numerically propagated under the full $N$-body model (corresponding to the blue line shown in the left panel).

It is observed from Fig. \ref{Fig10} that (a) both the trajectories propagated under the full $N$-body model and under the semi-secular model are in good agreement, (b) both the numerically propagated trajectories follow closely along the level curves arising in the phase portrait, (c) it is interesting to see that the resonant argument $\sigma$ of 2018 VO$_{137}$ can switch between libration and circulation in the long-term evolution (this behavior is in agreement with that predicted by the associated level curve arising in the phase portrait), (d) in the phase portrait there are four islands of the ZLK resonance and their centers are at $\omega = 0^{\circ}$, $\omega = 90^{\circ}$, $\omega = 180^{\circ}$ and $\omega = 270^{\circ}$, and (e) 2018 VO$_{137}$ is currently located inside the ZLK island centered at $\omega = 90^{\circ}$. It is noted that the trajectories produced from the semi-secular model are presented by averaged elements because short-term oscillations have been filtered out in the process of formulating the associated dynamical model. However, the trajectories produced from the full $N$-body model are described by osculating elements (with short-period oscillations).

Another two TNOs inside 2:5 resonance with Neptune are taken into account in Fig. \ref{Fig11}, where the level curves of adiabatic invariant $S$ (phase portraits) are plotted. The trajectories of these two TNOs are numerically propagated under the full $N$-body model and they are presented in blue lines. From Fig. \ref{Fig11}, it is observed that the numerically integrated trajectories follow closely along the level curves in the phase portraits.

For 2005 SD$_{278}$ (see the left panel of Fig. \ref{Fig11}), it is currently inside mean motion resonance with Neptune and, similar to 2018 VO$_{137}$ discussed in Fig. \ref{Fig10}, the resonant angle $\sigma$ could switch between libration and circulation. This is in agreement with the behavior predicted by the associated level curve of $S$ in the phase portrait. In the phase portrait associated with 2005 SD$_{278}$, there are four ZLK islands centered at $\omega = 0^{\circ}$, $\omega = 90^{\circ}$, $\omega = 180^{\circ}$ and $\omega=270^{\circ}$. In general, the islands centered at $\omega = 90^{\circ}$ and $\omega = 270^{\circ}$ are larger than the ones centered at $\omega = 0^{\circ}$ and $\omega = 180^{\circ}$ and the former ones hold lower eccentricities than the latter ones. Currently, 2005 SD$_{278}$ is inside the island centered at $\omega = 270^{\circ}$. Due to the ZLK resonance, the eccentricity of the TNO has a large range of variation. Naturally, coupled evolution between eccentricity and inclination indicates that the inclination should also have a large variation.

For 2015 PD$_{312}$ (see the right panel of Fig. \ref{Fig11}), there are four ZLK islands centered at $\omega = 0^{\circ}$, $\omega = 90^{\circ}$, $\omega = 180^{\circ}$ and $\omega=270^{\circ}$. The islands centered at $\omega = 0^{\circ}$ and $\omega = 180^{\circ}$ are located above the critical curve, meaning these two islands of ZLK resonance are inside the libration zones of MMR. In particular, 2015 PD$_{312}$ is currently inside the ZLK island centered at $\omega = 180^{\circ}$. It means that 2015 PD$_{312}$ is located inside the MMR and inside the ZLK resonance.

\begin{figure*}
\centering
\includegraphics[width=0.45\textwidth]{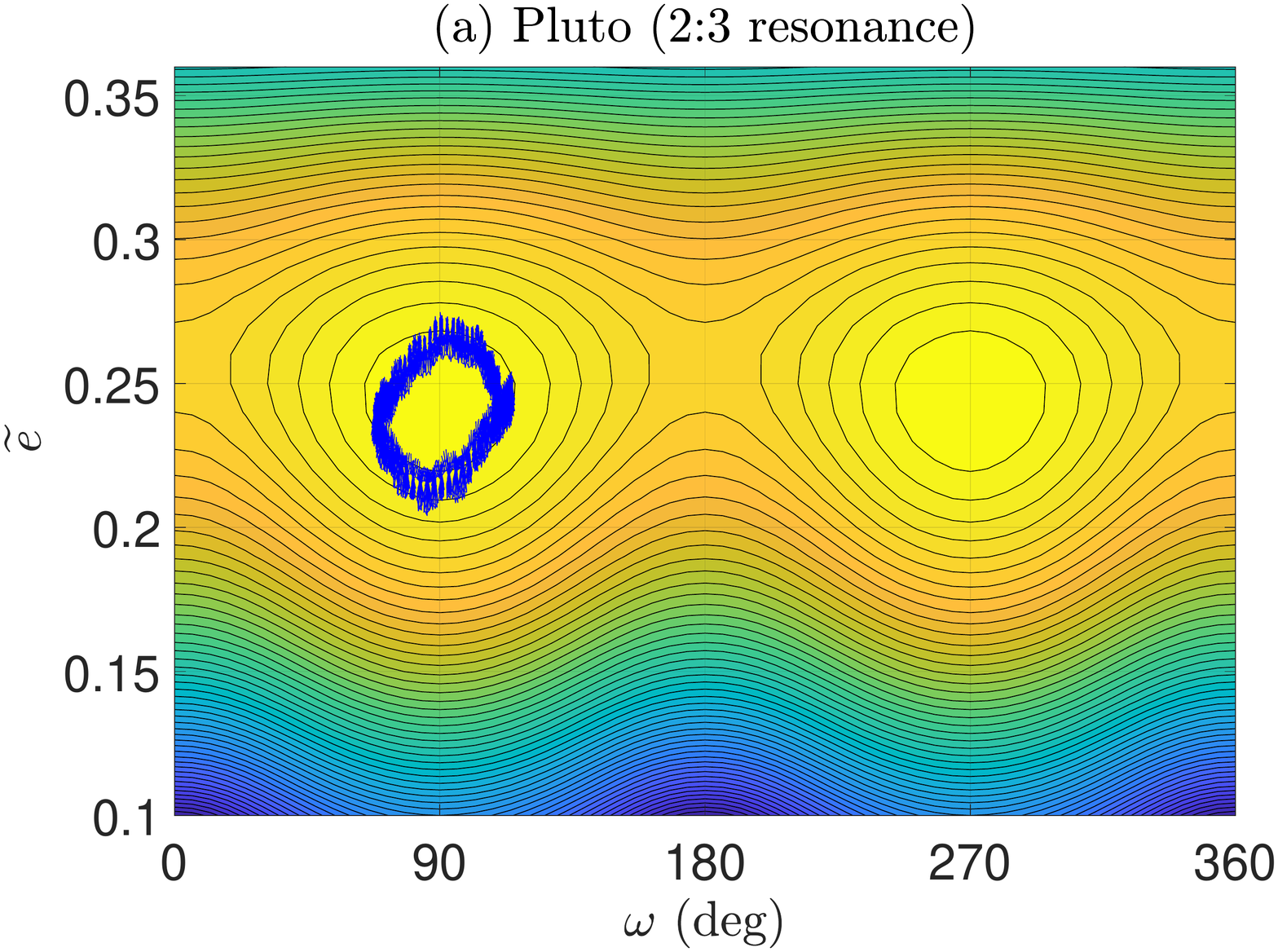}
\includegraphics[width=0.45\textwidth]{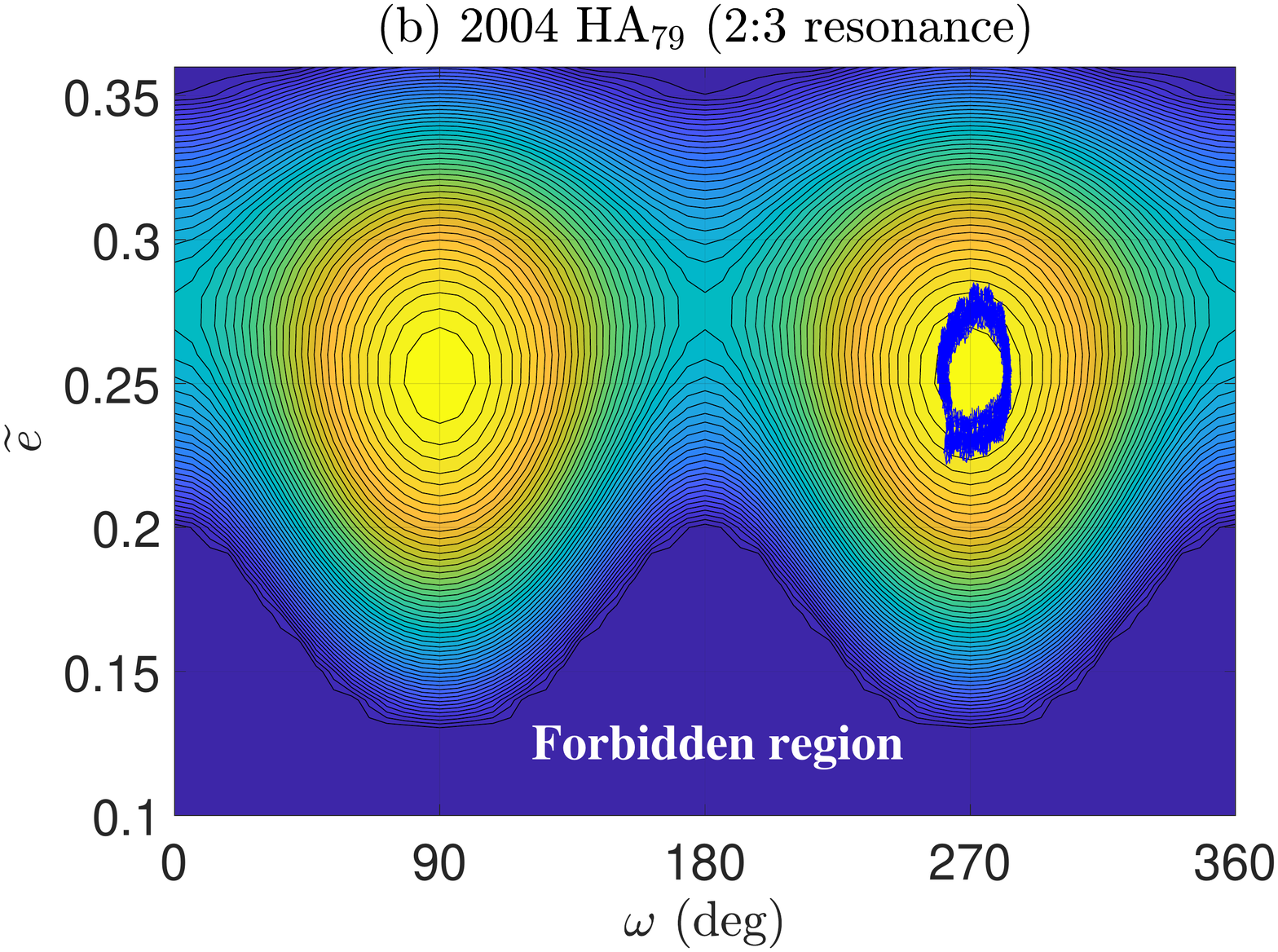}
\caption{Phase portraits together with the trajectories numerically propagated under the full $N$-body model for two TNOs inside 2:3 resonance with Neptune (Pluto and 2004 HA$_{79}$). These two TNOs are inside the ZLK resonance: one with $\omega$ librating around $90^{\circ}$ and the other one with $\omega$ librating around $270^{\circ}$).}
\label{Fig12}
\end{figure*}

Two TNOs inside 2:3 resonance with Neptune (Pluto and 2004 HA$_{79}$) are considered in Fig. \ref{Fig12}, where the phase portraits and numerically propagated trajectories under the full $N$-body model are presented. It is observed that (a) the numerical trajectories are in agreement with the level curves arising in the phase portraits and (b) Pluto is located inside the ZLK resonance with $\omega$ librating around $90^{\circ}$ and 2004 HA$_{79}$ is located inside the ZLK resonance with $\omega$ librating around $270^{\circ}$. For both TNOs considered here, the real location of libration center is coincident with the location predicted by the phase portrait. \citet{gladman2012resonant} showed that 2004 HA$_{79}$ is currently located inside ZLK resonance at $270^{\circ}$ with an amplitude of $30^{\circ}$, which is in agreement with our result. For TNOs trapped inside 2:3 MMR with Neptune, \citet{wan2007exploration} analytically formulated a secular dynamical model by fixing the critical argument of MMR at the libration center of $180^{\circ}$. An application to Pluto shows that the real libration center under the full $N$-body model is higher than the analytical prediction (see Figure 4 in their work), which is mainly due to the assumption adopted for the critical argument of MMR.

\begin{figure*}
\centering
\includegraphics[width=0.45\textwidth]{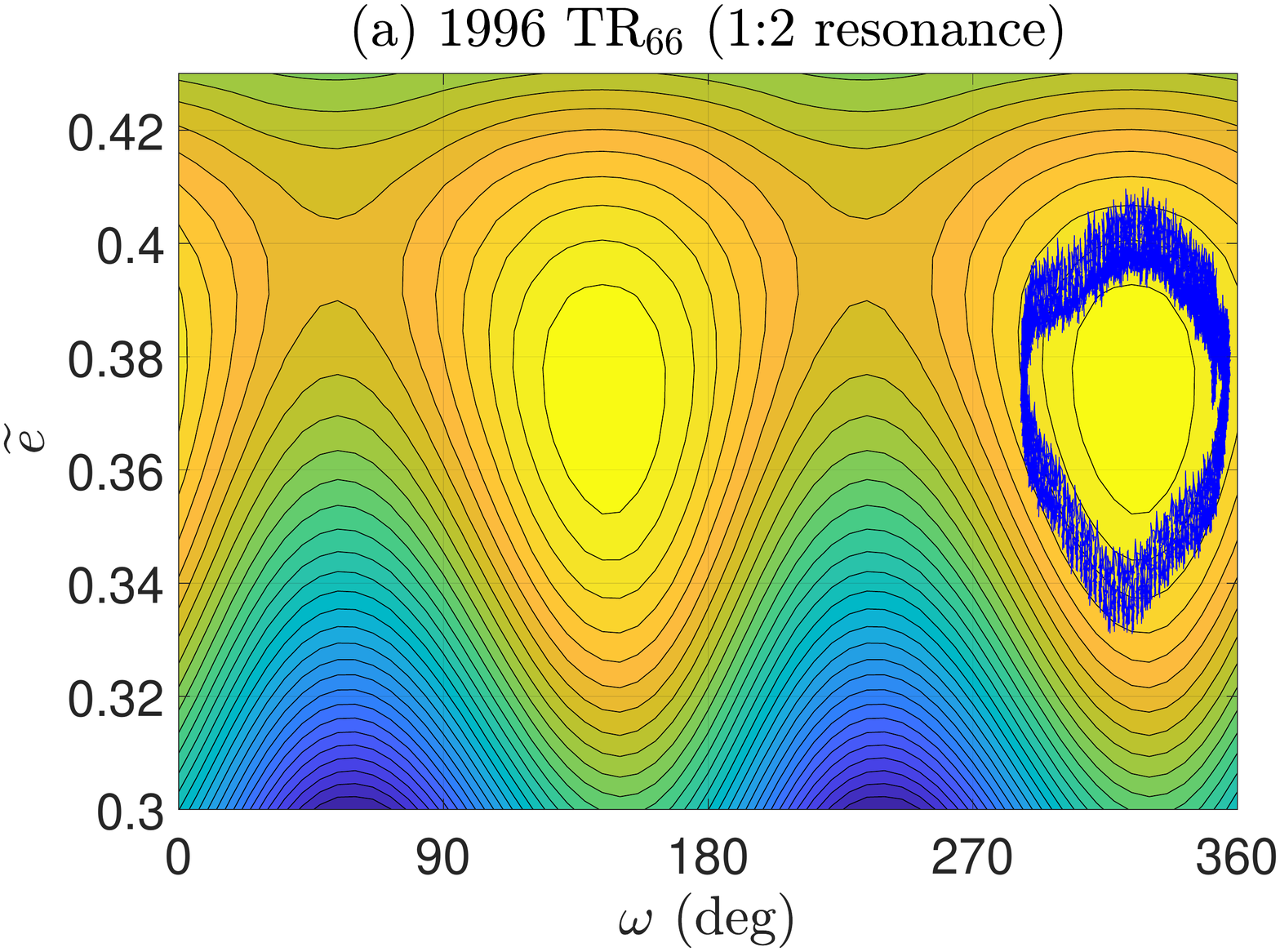}
\includegraphics[width=0.45\textwidth]{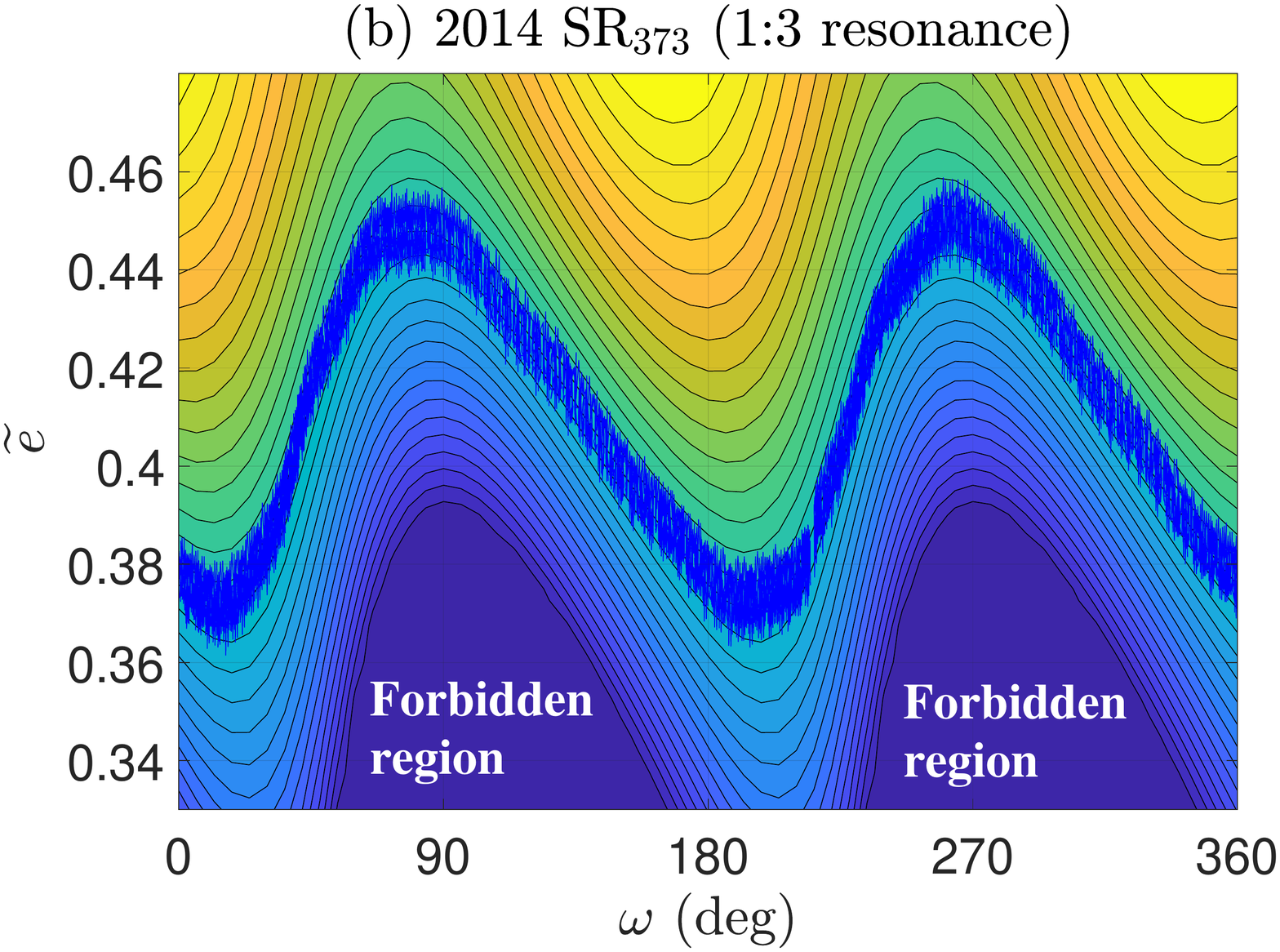}
\caption{Phase portraits together with the trajectories numerically propagated under the full $N$-body model for a TNO inside 1:2 resonance with Neptune (1996 TR$_{66}$) and a TNO inside 1:3 resonance with Neptune (2014 SR$_{373}$). 1996 TR$_{66}$ is inside the ZLK resonance, while 2014 SR$_{373}$ is outside the ZLK resonance.}
\label{Fig13}
\end{figure*}

Regarding the 1:n-type resonances with Neptune, we take the 1:2 and 1:3 resonances as examples. In particular, 1996 TR$_{66}$ inside 1:2 resonance and 2014 SR$_{373}$ inside 1:3 resonance are taken into account. Their phase portraits and trajectories numerically propagated under the full $N$-body model are reported in Fig. \ref{Fig13}. It is observed from Fig. \ref{Fig13} that (a) the numerical trajectories are in quite good agreement with the level curves appearing in the phase portraits and (b) 1996 TR$_{66}$ is inside the ZLK resonance while 2014 SR$_{373}$ is outside the ZLK resonance. In the left panel of Fig. \ref{Fig13}, it is observed that there are asymmetric centers of the ZLK islands. For the 1:n-type resonances, the centers of the ZLK islands are no longer at $\omega = 0^{\circ}, 90^{\circ}, 180^{\circ}$ and $270^{\circ}$. The asymmetric center of the ZLK resonance inside 1:n-type MMRs has been known in previous works \citep{kozai1985secular, gallardo2012survey, saillenfest2016long}.

\section{Summary and discussion}
\label{Sect6}

In this work, a semi-analytical one-degree-of-freedom model was formulated in order to explore the ZLK resonance of TNOs inside MMRs with Neptune. Firstly, we introduced a modified adiabatic invariant, denoted by $S$, which is equal to the absolute of the area enclosed by isolines of resonant Hamiltonian in the space $(\sigma, \Sigma)$. Compared to the traditional version, the adiabatic invariant adopted in this work is continuous around the dynamical separatrix. The continuous characteristic is very useful to describe those TNOs with switching behaviours between libration and circulation, because it doesn't need to match different phase portraits to describe a single TNO moving in different regions. Secondly, phase portraits are produced by plotting level curves of adiabatic invariant with given Hamiltonian, which can be used to predict long-term behaviours of TNOs. Compared to the conventional version of phase portraits (i.e., level curves of Hamiltonian with given adiabatic invariant), it requires smaller computational burden for producing new versions of phase portrait because the resonant Hamiltonian is an explicit function while the adiabatic invariant is an implicit function of the state variables $(\sigma, \Sigma, \omega, \tilde{e})$. Thirdly, in the new analytical model, it is possible to produce critical curves (the curves on which the Hamiltonian is equal to that of dynamical separatrix), which divide the entire phase space into different domains (with librations or circulations). Distributions of critical curve are very useful for us to predict dynamical behaviours of TNOs in the long-term evolution.

Analytical developments are applied to real TNOs inside MMRs with Neptune. In particular, three representative TNOs inside 2:5 resonance (2018 VO$_{137}$, 2005 SD$_{278}$ and 2015 PD$_{312}$), two representative TNOs inside 2:3 resonances (Pluto and 2004 HA$_{79}$), one TNO inside 1:2 resonance (1996 TR$_{66}$) and one TNO inside 1:3 resonance (2014 SR$_{373}$) are taken as examples. For each TNO considered, the trajectory is numerically propagated under the full $N$-body model and compared to the level curve arising in phase portraits. Good agreement is observed between numerical trajectories under the full $N$-body model and level curves arising in phase portraits, showing that our analytical developments are applicable for predicting long-term behaviours. In particular, level curves shown in phase portraits provide phase-space paths of eccentricity excitation, which is helpful to understand the origin and evolution of high-eccentricity TNOs. In phase portraits, different numbers of ZLK islands can be found. If there exist four ZLK islands, we can see that the islands centered at $180^{\circ}$ or $360^{\circ}$ occupy higher-eccentricity zones than the ones centered at $90^{\circ}$ or $270^{\circ}$. Regarding the non-1:n type of resonance, the ZLK centers are usually at symmetric locations of $\omega = 90^{\circ}$, $180^{\circ}$, $270^{\circ}$ and $360^{\circ}$, while for the 1:n-type resonances the ZLK centers are no longer at the symmetric locations, which is in agreement with previous works \citep{kozai1985secular, gallardo2012survey, saillenfest2016long}.

At last, some remarks are made here for explaining the slight differences between numerical results under the full $N$-body model and level curves in phase portraits (see Figs. \ref{Fig10}--\ref{Fig13}). From the dynamical viewpoint, there are several possible reasons leading to the difference. Firstly, the numerical trajectories propagated under the full $N$-body model are presented in osculating elements, so short-period oscillations can be observed in the long-term evolution. However, the level curves (or guiding trajectories in the secular model) stand for the evolution of averaged elements. Secondly, in the full $N$-body model, the secular resonances associated with Planets' precession rates as well as the ZLK resonance may together influence the long-term evolution of TNOs. However, in the semi-secular or secular models, there are no other secular resonances besides the ZLK resonance. Finally, in the full $N$-body model, the osculating orbits of giant planets are changed with time and their inclinations and eccentricities are not equal to zero (although they are small). In the semi-secular or secular dynamical models, the orbits of giant planets are fixed and they are assumed as coplanar and circular.


\bibliography{mybib}{}
\bibliographystyle{aasjournal}



\end{document}